\documentclass[cernpreprint, atlasdraft=false, coverpage=false, texlive=2023, UKenglish, texmf, orcidlogo]{atlasdoc}
 
\usepackage[full]{atlaspackage}
\usepackage{atlasbiblatex}
\usepackage[normalem]{ulem}
 
\usepackage[particle,process,jetetmiss,snippets,other]{atlasphysics}
 
\graphicspath{{logos/}{figures/}}
 
\usepackage{ANA-TOPQ-2020-27-PAPER-defs}

\AtlasTitle{Search for leptonic charge asymmetry in $t\bar{t}W$ production in final states with three leptons at $\sqrt{s} = 13$~TeV}
 
\AtlasVersion{2.8}

\AtlasAbstract{
A search for the leptonic charge asymmetry~($A_\text{c}^{\ell}$) of top-quark--antiquark pair production in association with a $W$~boson~($t\bar{t}W$) is presented. The search is
performed using final states with exactly three charged light leptons~(electrons or muons) and is based on $\sqrt{s} = 13$~TeV proton--proton collision data collected with the
ATLAS detector at the Large Hadron Collider at CERN during the years 2015--2018, corresponding to an integrated luminosity of 139~fb$^{-1}$. A profile-likelihood fit to the
event yields in multiple regions corresponding to positive and negative differences between the pseudorapidities of the charged leptons from top-quark and top-antiquark decays
is used to extract the charge asymmetry. At reconstruction level, the asymmetry is found to be $-0.12 \pm 0.14$~(stat.)~$\pm \, 0.05$~(syst.). An unfolding procedure is applied to
convert the result at reconstruction level into a charge-asymmetry value in a fiducial volume at particle level with the result of $-0.11 \pm 0.17$~(stat.)~$\pm \, 0.05$~(syst.).
The Standard Model expectations for these two observables are calculated using Monte Carlo simulations with next-to-leading-order plus parton shower precision in quantum chromodynamics
and including next-to-leading-order electroweak corrections. They are $-0.084 \, ^{+0.005}_{-0.003}$~(scale) $\pm\, 0.006$~(MC stat.) and $-0.063 \, ^{+0.007}_{-0.004}$~(scale)
$\pm\, 0.004$~(MC stat.) respectively, and in agreement with the measurements.
}
 
\AtlasRefCode{TOPQ-2020-27}
 
\PreprintIdNumber{CERN-EP-2022-249}
 
\AtlasDate{\today}

\AtlasJournal{JHEP}
\AtlasJournalRef{\JHEP 07 (2023) 033}
\AtlasDOI{10.1007/JHEP07(2023)033}

\hypersetup{pdftitle={ATLAS document},pdfauthor={The ATLAS Collaboration}}
 
\begin{document}
 
\maketitle
 
\section{Introduction}

The production of a top-quark--antiquark~(\ttbar) pair in association with a \Wboson~boson, commonly referred to as \ttW, is a rare process in the Standard Model~(SM) that can
be produced at the Large Hadron Collider~(LHC). State-of-the-art cross-section calculations for the \ttW process are especially complex, as large corrections arise from higher
powers of both the strong and weak couplings~\cite{Broggio:2019ewu}. Thus, measurements of the \ttW process represent a sensitive test of the predictions of quantum chromodynamics~(QCD)
and the electroweak~(EW) sector of the SM, as well as their interplay. Both the inclusive and differential cross-section measurements are very relevant, as they can provide indirect
hints of new physics beyond the SM~(BSM) in scenarios where at least one of the SM couplings is modified~\cite{Grzadkowski:2010es}. Furthermore, the \ttW process can be one of the main backgrounds
in searches for BSM phenomena, such as supersymmetric squark or gluino production or vector-like quarks~\cite{SUSY-2018-09,EXOT-2016-16}. It also represents an irreducible background
in many measurements of SM processes such as \ttbar production in association with a Higgs boson~(\ttH) or the production of four top quarks~(\tttt)~\cite{HIGG-2017-02,TOPQ-2020-10}.
The inclusive cross-section of \ttW production has been measured by both the ATLAS and CMS collaborations at $\rts = \SI{13}{\TeV}$ using partial and full LHC \RunTwo
datasets~\cite{TOPQ-2016-11,CMS:2022tkv}, respectively.
 
\begin{figure}[!htpb]
\centering
\includegraphics[width=0.95\textwidth]{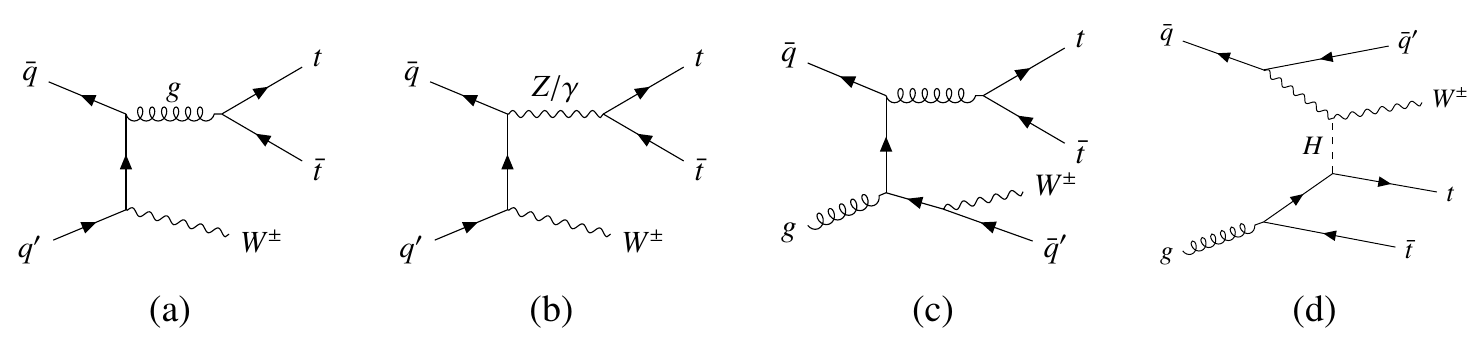}
\caption{Examples of Feynman diagrams of \ttW production at (a,b)~LO and (c,d)~NLO with one extra parton. The diagrams show (a,c)~QCD and (b,d)~EW \ttW production.}
\label{fig:ttwproc}
\end{figure}

Illustrative Feynman diagrams contributing to \ttW production at leading order~(LO) and next-to-leading order~(NLO) for both QCD and EW production are shown
in Figure~\ref{fig:ttwproc} where $q^\prime$ indicates a quark of different flavour from that of the other initial-state quark. At LO, only the $\qqbar^\prime$ initial state
is present~(Figure~\ref{fig:ttwproc}~a,b). At NLO, the quark--gluon~($qg$) channels open up~(Figure~\ref{fig:ttwproc}~c,d), whereas gluon--gluon~($gg$) fusion production does
not contribute until next-to-next-to-leading-order~(NNLO) corrections are included.
 
In \ttbar production, the top quark~(top antiquark) is preferentially produced in the direction of the incoming quark~(antiquark). This is due to the interference effects between
amplitudes in the \qqbar initial state and results in a difference in the rapidity distribution between top quarks and top antiquarks.\footnote{\AtlasCoordFootnote} In
proton--proton~(\pp) collisions at the LHC, this production asymmetry results in a central--forward rapidity charge asymmetry as top quarks~(antiquarks) are produced with
more forward~(central) rapidities. Given that \ttbar production at the LHC is dominated by the charge-symmetric $gg$ initial state, such asymmetry is a subtle~(order of 1\%)
effect. This is different from the situation at the Tevatron collider ($p\bar{p}$ collisions), where a forward--backward asymmetry can be defined with respect to the proton beam, and \qqbar collisions dominate over $gg$, yielding a more sizable signal (10\%)~\cite{CDF:2017cvy}.

The top-quark-based rapidity charge asymmetry~(\tca) is defined by
\begin{equation}
\tca = \frac{N \,(\Delta y^{t} > 0) - N \,(\Delta y^{t} < 0)}{N \,(\Delta y^{t} > 0) + N \,(\Delta y^{t} < 0)},
\label{eq:lca_formula1}
\end{equation}
where $\Delta y^{t} = |y_{t}| - |y_{\tbar}|$ is the difference between the absolute rapidities of the top quark~($|y_{t}|$) and top antiquark~($|y_{\tbar}|$), respectively.
 
In \ttW production, the relative dominance of the $\qqbar^\prime$ initial state leads to a larger rapidity charge asymmetry than in \ttbar production~\cite{Maltoni:2014zpa,Bevilacqua:2020srb}.
Furthermore, the \Wboson~boson present in \ttW production is typically radiated from the initial $\qqbar^\prime$ state and, therefore, serves as a polarity filter of the initial $\qqbar^\prime$
state and in turn the final \ttbar state. This polarisation further enhances the asymmetry between the decay products of the top quarks and top antiquarks. The prospects for experimental
observation of these asymmetries are greatest in the case of the charged leptons originating from the top-quark~(antiquark) decays. This is due to the precision with which the lepton
kinematics can be reconstructed and the power with which reducible background processes can be suppressed. The leptonic charge asymmetry~(\lca), in the following just referred to as
\enquote{charge asymmetry}, is defined analogously to Eq.~(\ref{eq:lca_formula1}), but based on the pseudorapidities of the leptons from the top-quark and top-antiquark decays:
\begin{equation}
\lca = \frac{N \,(\Delta \eta^{\ell} > 0) - N \,(\Delta \eta^{\ell} < 0)}{N \,(\Delta \eta^{\ell} > 0) + N \,(\Delta \eta^{\ell} < 0)},
\label{eq:lca_formula2}
\end{equation}
where $\Delta \eta^{\ell} = |\eta_{\ell}| - |\eta_{\bar{\ell}}|$ is the difference between the absolute pseudorapidities of the leptons that originate from the top quark~($|\eta_{\ell}|$) and
top antiquark~($|\eta_{\bar{\ell}}|$), respectively.
 
Reference~\cite{Maltoni:2014zpa} gives a comparison of NLO QCD matrix elements~(MEs) matched to parton shower~(PS) calculations of the top-quark-based and leptonic charge asymmetries
of \ttbar and \ttW production in the full phase space at $\rts = \SI{13}{\TeV}$. The charge asymmetry for \ttW is larger than for \ttbar production at the expense of a smaller
cross-section for the process. In addition to being sensitive to BSM physics, such as axigluons and Standard Model Effective Field Theory~(SMEFT) scenarios corresponding to four-fermion
operators~(examples given in Refs.~\cite{Maltoni:2014zpa} and~\cite{Brivio:2017vri}), charge asymmetry measurements have the potential to discriminate between new physics signals with
different chiral structures that would be indistinguishable using only cross-section observables.
 
At the Tevatron, forward--backward asymmetries in \ttbar production have been measured, with results found to be in agreement with SM calculations that include higher-order
corrections~\cite{CDF:2011xdt,CDF:2017cvy}. The ATLAS and CMS collaborations performed measurements of the top-quark-based charge asymmetry for \ttbar production.
A combination of these ATLAS and CMS results at $\rts = \SI{7}{\TeV}$ and \SI{8}{\TeV} for the top-quark-based charge asymmetry is reported in Ref.~\cite{TOPQ-2016-16} and updated
measurements have been published by ATLAS and CMS using $\rts = \SI{13}{\TeV}$ data~\cite{TOPQ-2020-06,CMS:2022ged}. The measurements reported by CMS in Ref.~\cite{CMS-TOP-17-014}
include an extraction of the leptonic charge asymmetry for \ttbar production in a particle-level fiducial volume. A measurement of the top-quark-based charge asymmetry for \ttbar
production in association with a photon has been reported by ATLAS in Ref.~\cite{ATLAS:2022wec}. None of these measurements show significant deviations from the SM expectations.
In Ref.~\cite{Bevilacqua:2020srb}, NLO QCD calculations of \lca have been performed including top-quark off-shell effects, which also include the impact of different renormalisation
and factorisation scale choices on \lca in the multi-lepton channel at the LHC at $\rts = \SI{13}{\TeV}$.\footnote{These results are given in terms of the rapidities of the
leptons~($A_{\text{c},y}^{\ell}$) and not the pseudorapidities.}

This paper presents a search for the leptonic charge asymmetry in \ttW production using \pp collision data at $\rts = \SI{13}{\TeV}$ in the trilepton~(\lll) channel
with the full \RunTwo data sample, corresponding to an integrated luminosity of \SI{\lumi}{\ifb}. The paper is organised as follows. Section~\ref{sec:detector} provides a brief description
of the ATLAS detector. In Section~\ref{sec:samples}, the data sample as well as the simulated signal and background processes are discussed. The reconstructed particle candidates
are defined in Section~\ref{sec:objects}. Section~\ref{sec:regions} gives an overview of the event selection and of the definitions of the control and signal regions. The algorithm
used to identify reconstructed leptons originating from top quarks~(antiquarks) is explained in Section~\ref{sec:leptontop}. In Section~\ref{sec:systematics}, the sources of systematic
uncertainties that affect the search are discussed. The result for the charge asymmetry measurement at reconstruction level is presented in Section~\ref{sec:results}. The unfolding
procedure and the extraction of the charge asymmetry at particle level are presented in Section~\ref{sec:unfolding}. In Section~\ref{sec:conclusions}, the conclusions are drawn.


\section{The ATLAS detector}
\label{sec:detector}

The ATLAS detector~\cite{PERF-2007-01} at the LHC covers nearly the entire solid angle around the collision point. It consists of an inner
tracking detector surrounded by a thin superconducting solenoid, electromagnetic and hadronic calorimeters, and a muon spectrometer incorporating
three sets of large superconducting toroidal magnets, each consisting of eight separate coils. The inner-detector system is immersed in a
\SI{2}{\tesla} axial magnetic field and provides charged-particle tracking in the range $\abseta < 2.5$.
 
The high-granularity silicon pixel detector covers the vertex region and typically provides four measurements per track, with the first hit typically
being  detected in the insertable B-layer installed before \RunTwo~\cite{ATLAS-TDR-2010-19,PIX-2018-001}. It is followed by the silicon microstrip
tracker, which usually provides eight measurements per track. These silicon detectors are complemented by the transition radiation tracker~(TRT),
which enables radially extended track reconstruction up to $\abseta = 2.5$. The TRT also provides electron identification information based on the
fraction of hits above a higher energy-deposit threshold corresponding to transition radiation. Typically, around \num{30} TRT hits are measured in
total per track.
 
The calorimeter system covers the pseudorapidity range $\abseta < 4.9$. In the region $\abseta < 3.2$, electromagnetic calorimetry is provided
by barrel and endcap high-granularity lead/liquid-argon~(LAr) calorimeters, with an additional thin LAr presampler covering $\abseta < 1.8$ to correct
for energy loss in material upstream of the calorimeters. Hadronic calorimetry is provided by the steel/scintillator-tile calorimeter, segmented into
three barrel structures with $\abseta < 1.7$, and two copper/LAr hadronic endcap calorimeters. The solid angle coverage is extended with forward
copper/LAr and tungsten/LAr calorimeter modules optimised for electromagnetic and hadronic measurements respectively.
 
The muon spectrometer comprises separate trigger and high-precision tracking chambers measuring the deflection of muons in a magnetic field generated
by the superconducting air-core toroids. The field integral of the toroids ranges between \num{2.0} and \SI{6.0}{Tm} across most of the detector. A set of
precision chambers covers the region $\abseta < 2.7$ with three layers of monitored drift tubes, complemented by cathode-strip chambers in the forward
region, where the background rates are highest. The muon trigger system covers the range $\abseta < 2.4$ with resistive-plate chambers in the barrel,
and thin-gap chambers in the endcap regions.
 
Relevant events are selected to be recorded by the first-level trigger system implemented in custom hardware, followed by selections made by algorithms
implemented in software in the high-level trigger~\cite{TRIG-2016-01}. The first-level trigger accepts events from the \SI{40}{\MHz} bunch crossings at a
rate below \SI{100}{\kHz}, which the high-level trigger reduces to record events to disk at about \SI{1}{\kHz}.
 
An extensive software suite~\cite{ATL-SOFT-PUB-2021-001} is used in data simulation, in the reconstruction and analysis of real and simulated data, in detector operations,
and in the trigger and data acquisition systems of the experiment.


\section{Data and simulated event samples}
\label{sec:samples}

The analysis is performed on data from \pp collisions at $\rts = \SI{13}{\TeV}$ delivered by the LHC and recorded by the ATLAS detector in the years 2015--2018. The
bunch spacing for this data-taking period was \SI{25}{ns} with a typical number of \pp interactions per bunch crossing~(\enquote{pile-up}) that varies by year
and LHC beam conditions and was in the range from \num{10} to \num{70} for almost all events. After requirements on the stability of the beams, the operational
status of all ATLAS detector components, and the quality of the recorded data, the total integrated luminosity of the data sample corresponds to \SI{\lumi}{\ifb} with an uncertainty of 1.7\%.
This value is derived from the calibration of the luminosity scale using $x$--$y$ beam-separation scans, following a methodology similar to that detailed in
Ref.~\cite{ATLAS-CONF-2019-021}, and using the LUCID-2 detector~\cite{Avoni:2018} for the baseline luminosity measurements.
 
Simulated Monte Carlo~(MC) samples are used to model the contributions from the various SM processes. The MC generators used for the hard-scattering, as well as
the PS, underlying event and hadronisation, are explained in the following. For some processes, in addition to the nominal simulation, alternative MC samples are
available that are used to evaluate the effects of different MC modelling uncertainties~(see Section~\ref{subsec:sys-theo}). All MC samples were generated using
a \SI{25}{ns} bunch-spacing configuration.
 
The effect of pile-up was modelled by overlaying the hard-scattering event with simulated minimum-bias events generated with \PYTHIA[8.186]~\cite{Sjostrand:2007gs}
using the \NNPDF[2.3lo]~\cite{Ball:2012cx} set of parton distribution functions~(PDFs) and the A3 set of tuned MC parameters~\cite{ATL-PHYS-PUB-2016-017}. Separate MC
production campaigns were used to model the different pile-up distributions observed in data for the years 2015/16, 2017 and 2018. The simulated samples were reweighted to
reproduce the observed distribution of the average number of collisions per bunch crossing. The simulation of detector effects was performed with either a full ATLAS detector
simulation based on the \GEANT~\cite{Agostinelli:2002hh} framework or a fast simulation~(\textsc{AtlFast-II}) using a parameterisation of the performance of the electromagnetic
and hadronic calorimeters and \GEANT for the other detector components~\cite{SOFT-2010-01}.

The signal process~(\ttW) was simulated at NLO precision in QCD with \SHERPA[2.2.10]~\cite{Bothmann:2019yzt} and the \NNPDF[3.0nnlo] PDF set~\cite{Ball:2014uwa}.
In this set-up, multiple MEs were matched and merged with the \SHERPA PS model based on the Catani--Seymour dipole factorisation scheme~\cite{Gleisberg:2008fv,Schumann:2007mg}.
The virtual QCD corrections for MEs at NLO accuracy were provided by the \OPENLOOPS library~\cite{Cascioli:2011va,Denner:2016kdg}. Up to one additional parton was included in
the NLO ME, and two, three or four additional partons were included at LO in QCD. The merging scale parameter, which sets a threshold to determine what part of the phase-space
is filled by the PS or the ME generator, was set to an energy of \SI{30}{\GeV}. Additional partons beyond ME-level accuracy and below the merging scale threshold were therefore described
by the PS. The choice of renormalisation and factorisation scales was $\mu_R = \mu_F =
H_\textrm{T}$/2, where $H_\textrm{T}$ is defined as the scalar sum of
the transverse masses $\sqrt{p_\textrm{T}^2+m^2}$ of all final state
particles. The masses of the top quark and the \Wboson~boson were set to \SI{172.5}{\GeV} and \SI{80.4}{\GeV}, respectively~\cite{Workman:2022ynf}. In addition to the nominal prediction
at NLO in QCD~(order of $\alpha \alphas^{3}$),\footnote{$\alpha$ and \alphas denote the EW and strong coupling constants, respectively.} higher-order corrections related to EW \ttW
contributions were also added as part of the signal definition. The $\alpha^{3}$ and $\alpha^{2} \alphas^{2}$ corrections were added through MC event weights derived using the virtual
additive corrections in the formalism described in Ref.~\cite{Kallweit:2015dum}.
 
An alternative \ttW sample uses \MGNLO[2.9.3]~\cite{Alwall:2011uj-c,Alwall:2014hca}~(in the following denoted by \MG) for the ME and was
interfaced to \PYTHIA[8.245]~\cite{Sjostrand:2014zea} for the PS, underlying event and hadronisation modelling. This sample was generated with the \FXFX
algorithm~\cite{Frederix:2012ps} with up to one additional parton at NLO accuracy and up to two additional partons at LO accuracy in QCD. The expected accuracy of this
sample is similar to that of the nominal \SHERPA[2.2.10] sample. This multi-leg configuration makes use of complex functional forms for the renormalisation and
factorisation scales that are chosen dynamically and depend on the kinematics of the event after the merging of the core process with the additional
partons following the \FXFX merging prescription~\cite{Frederix:2012ps,Catani:2001cc,Hamilton:2012np}. They depend on the phase-space configuration and are
related to the clustering scales of the additional partons and on the core process. The merging scale parameter was set to \SI{30}{\GeV}. The sample was simulated using
the \NNPDF[3.0nlo] PDF set and the A14 set of tuned MC parameters~\cite{ATL-PHYS-PUB-2014-021}, henceforth referred to as the \enquote{A14 MC tune}. Top-quark decays
were simulated at LO using the \MADSPIN program~\cite{Frixione:2007zp,Artoisenet:2012st}. Further alternative \ttW samples were simulated with the \POWHEG~\cite{Frixione:2007vw}
generator providing ME calculations at NLO in \alphas with the \NNPDF[3.0nlo] PDF set and the A14 MC tune. These \POWHEG \ttW samples were interfaced either with
\PYTHIA[8.245] or with \HERWIG[7.2.1]~\cite{Corcella:2000bw,Bahr:2008pv,Bellm:2019zci} for the simulation of the PS, underlying event and hadronisation. All the alternative
\ttW samples were normalised to the same cross-section as the nominal \SHERPA[2.2.10] sample in order not to be sensitive to overall normalisation differences when comparing
the two simulations.
 
The \ttW EW corrections at the order of $\alpha^{3} \alphas$ were simulated by an independent \SHERPA[2.2.10] sample, produced at LO in QCD with the same configuration as
the nominal signal sample. Since the charge asymmetry from these processes is negligible compared to the nominal signal contribution, this MC sample is treated as a
background in the analysis.
 
The production of a \ttbar pair in association with a \Zboson~boson~(\ttZ) was simulated at NLO precision with the \MG[2.8.1] generator for the ME and \PYTHIA[8.244] for the PS, underlying
event and hadronisation, together with the \NNPDF[3.0nlo] PDF set and the A14 MC tune. The mass of the \Zboson~boson was set to \SI{91.2}{\GeV}~\cite{Workman:2022ynf}. The $\ttbar\gam^{*}$
contribution and $Z/\gam^{*}$ interference effects were taken into account, with the samples including events with dilepton invariant masses~($m_{\ell\ell}$) down to \SI{1}{\GeV},
where $\ell$ is an electron or muon. Additional \ttZ samples using \MG[2.8.1] for the ME, but \HERWIG[7.2.1] for the PS along with the \HERWIG standard set of tuned parameters and
the \NNPDF[3.0nlo] PDF set were used for the evaluation of systematic uncertainties associated with the PS and hadronisation. Further alternative \ttZ samples with the same settings
as the nominal samples, but using the A14 eigentune variation Var3c~\cite{ATL-PHYS-PUB-2014-021}, are used to evaluate the uncertainty associated with the initial-state radiation~(ISR).
Similarly to \ttW, the alternative \ttZ samples were normalised to the same cross-section as the nominal sample.
 
The production of \ttbar, \ttH and \tW events was simulated at NLO with the \POWHEG generator for the ME, together with the \NNPDF[3.0nlo] PDF set and the A14 MC tune. The $h_{\text{damp}}$
parameter, which controls the matching in \POWHEG and regulates the high-\pt radiation against which the \ttbar system recoils, was set to \num{1.5} times the nominal top-quark mass~\cite{ATL-PHYS-PUB-2016-020}.
The events were interfaced to \PYTHIA[8.230] for the PS, underlying event and hadronisation. The \ttbar cross-section was normalised to next-to-next-to-leading-logarithmic order~(NNLL)
in QCD, including the resummation of NNLL soft-gluon terms~(NNLO~+~NNLL)~\cite{Melnikov:2011ta,Czakon:2012pz}. The \ttH samples were normalised to NLO~(QCD and EW) using the calculations
documented in Ref.~\cite{deFlorian:2016spz}. The \tW sample was normalised to NLO in QCD including NNLL soft-gluon corrections~\cite{Kidonakis:2010ux}. An alternative \ttbar simulation
was used with the same set-up for the ME, but the events were interfaced to \HERWIG[7.1.3]~\cite{Bellm:2017jjp} for the PS, underlying event and hadronisation modelling. The \HERWIG
standard set of tuned parameters and the \NNPDF[3.0nlo] PDF set were used. Alternative \ttH samples were used, where either the ME generator~(\MG[2.6.0]) or the PS algorithm~(\HERWIG[7.2.1])
was changed with respect to the nominal \ttH simulation.
 
A MC sample featuring the production of \ttbar events in association with photons~(\tty) was simulated at LO in QCD with \MG[2.3.3] and interfaced to \PYTHIA[8.212], together
with the \NNPDF[3.0nlo] PDF set and the A14 MC tune. This sample is, however, only used to assign an extra uncertainty to additional photon radiation in the nominal \ttbar
prediction. Details of this procedure can be found in Section~\ref{subsec:sys-theo}.
 
The production of a single top quark~(or antiquark) in association with a \Zboson~boson and one extra parton~(\tZq) was simulated using the \MG[2.3.3] generator at NLO
with the \NNPDF[3.0nnlo] PDF set. The events were interfaced to \PYTHIA[8.245] using the A14 MC tune. The \tZq simulation also includes off-shell \Zboson~boson decays into dilepton
pairs with invariant masses in the range $m_{\ell\ell} > \SI{5}{\GeV}$. Single top quark~(antiquark) production in association with both a \Wboson and a \Zboson boson~(\tWZ) was simulated
at NLO with \MG[2.2.2] and the \NNPDF[3.0nnlo] PDF set, using \PYTHIA[8.235] for the PS simulation. The interference between \ttZ and \tWZ was removed following a diagram-removal~(DR)
approach referred to as the \enquote{DR1 scheme}~\cite{Demartin:2016axk}.
 
The MC samples featuring \Zjets production were simulated at NLO with the \POWHEG generator for the ME and interfaced to \PYTHIA[8.186] for the PS. The AZNLO~\cite{STDM-2012-23}
set of tuned parameters and the \NNPDF[3.0nnlo] PDF set were used. An alternative \Zjets simulation was done with the \SHERPA[2.2.11] generator where the default \SHERPA PS set-up
was used along with the \NNPDF[3.0nnlo] PDF set. The \Zjets samples feature events with $m_{\ell\ell}$ down to \SI{10}{\GeV}. The sample cross-sections were normalised to NNLO
predictions~\cite{Melnikov:2006kv}. The \POWHEG~+~\PYTHIA[8] sample used \PHOTOS~\cite{Golonka:2005pn} for final-state radiation~(FSR).
 
For the simulation of \Zboson~boson production in association with a photon~(\Zy), \SHERPA[2.2.11] was used with the \NNPDF[3.0nnlo] PDF set. The events were simulated
at NLO precision.
 
Diboson processes featuring the production of three charged leptons and one neutrino or four charged leptons~(denoted by \WZjets or \ZZjets, respectively) were simulated
using the \SHERPA[2.2.2] generator, with a similar set-up to that described for \Zjets. Events with up to one extra parton were simulated at NLO, and with two or three partons at LO precision.
MC samples featuring Higgs boson production in association with a \Wboson or \Zboson~boson~(\HWZ) were generated at NLO using \POWHEG interfaced to \PYTHIA[8.230/8.235] for the PS, together
with the \NNPDF[3.0nlo] PDF set and the AZNLO MC tune.
 
The production of four top quarks~(\tttt) was modelled at NLO with \SHERPA[2.2.11] together with the \NNPDF[3.0nnlo] PDF set. The production of three top quarks~(\ttt)
and the production of a \ttbar pair with two \Wboson~bosons~(\ttWW) were simulated at LO using \MG[2.2.2] interfaced to \PYTHIA[8.186] with the A14 MC tune and the
\NNPDF[2.3lo] PDF set. Fully leptonically decaying triboson processes~($WWW$, $WWZ$, $WZZ$ and $ZZZ$) with up to six leptons in the final states were simulated with \SHERPA[2.2.2]
and the \NNPDF[3.0nlo] PDF set. Final states with no additional partons were calculated at NLO, whereas final states with one, two or three additional partons were calculated at LO.
 
For all MC samples except the \SHERPA ones, the decays of $b$- and $c$-hadrons were simulated using the \EVTGEN[1.2.0] program~\cite{Lange:2001uf}.


\section{Event reconstruction}
\label{sec:objects}

Electron candidates are reconstructed from clusters of energy deposits in the electromagnetic calorimeter that are matched to tracks in the inner detector.
They are required to satisfy $\pt > \SI{10}{\GeV}$, $\abseta < 2.47$ and need to pass a \enquote{Tight} working point~(WP), defined by a likelihood-based
electron identification~(ID) requirement~\cite{EGAM-2018-01}. To reject non-prompt electrons, the reconstructed track associated with the electron
must satisfy the requirements $|\zzsth| < \SI{0.5}{mm}$ and $|\dzero| / \sigma(\dzero) < 5$, where $z_{0}$ describes the longitudinal impact parameter relative
to the reconstructed primary vertex,\footnote{The primary vertex is defined as the vertex~(at least two associated tracks with $\pt > \SI{500}{\MeV}$) with the highest
scalar sum of the squared transverse momenta of the associated tracks.} \dzero is the transverse impact parameter relative to the beam axis, and $\sigma(\dzero)$ is the
uncertainty in \dzero. Electron candidates are excluded if their calorimeter energy clusters lie within $1.37 < \abseta < 1.52$, the transition region between the barrel
and the endcap of the electromagnetic calorimeter.
 
Additional requirements are applied to the electron candidates to suppress the contribution of electrons originating from converted photons~(\gam-conversions). Electrons
can be identified as internal- or material-conversion candidates by checking for additional tracks close to the calorimeter energy clusters associated with the electrons and the
existence of conversion vertices. Electrons that are identified as either internal- or material-conversion candidates are rejected. These requirements are referred to in the following
as \enquote{$e/\gam$ ambiguity requirements}.
 
Furthermore, the electrons selected for the signal regions~(SRs) of the analysis have to satisfy an isolation requirement. An isolation WP is defined using a
multivariate likelihood discriminant that combines electromagnetic shower shapes and track information from the inner detector to distinguish prompt electrons from non-prompt/fake
electrons originating from hadronic jets, \gam-conversions and heavy-flavour hadron decays. The electrons satisfying the isolation requirements are henceforth referred to as \emph{tight}
electrons, whereas \emph{loose} electrons are defined by the conditions listed in the previous paragraph and do not need to satisfy the isolation requirement.
 
Muon candidates have to satisfy $\pt > \SI{10}{\GeV}$, $\abseta < 2.5$ and an ID selection, corresponding to a \enquote{Medium} WP. This sets requirements on the number of hits in
the different inner detector and muon spectrometer subsystems, and on the muon \pt and charge-to-momentum ratio variables~\cite{PERF-2015-10,MUON-2018-03}. If a muon has
insufficient momentum resolution, the entire event is removed. Requirements on the impact parameters of the reconstructed track associated with the muon candidate are applied to reject
non-prompt muons. The track is required to have $|\zzsth| < \SI{0.5}{mm}$ and $|\dzero| / \sigma(\dzero) < 3$.
 
As with electrons, an isolation requirement is applied to the muons used in the SRs, using the same approach based on a multivariate likelihood discriminant to distinguish
prompt muons from non-prompt/fake muons. Muons are referred to as \emph{tight} or \emph{loose}, depending on whether or not they satisfy this isolation requirement in addition to
the criteria listed above.
 
Jets are reconstructed using the \antikt jet algorithm~\cite{Cacciari:2008gp} on particle-flow objects~\cite{PERF-2015-09} with the radius parameter set to $R = 0.4$, as
implemented in the \Fastjet package~\cite{Fastjet}. The \antikt algorithm is used to reconstruct jets with a four-momentum recombination scheme, using the particle-flow objects
as inputs. The jet calibration is performed using a standard procedure that corrects the jet energy to match, on average, the particle-level jet energy in simulation
and applies an in situ correction for data~\cite{JETM-2018-05}. To suppress jets from additional \pp interactions within the same bunch crossing, a \enquote{jet vertex
tagger}~(\JVT)~\cite{ATLAS-CONF-2014-018,PERF-2014-03} is applied to select jets. The jets are only kept if they have $\pt > \SI{20}{\GeV}$ and are in a pseudorapidity
range of $\abseta < 2.5$. In addition, $\JVT > 0.2$ is required for jets with $\pt < \SI{60}{\GeV}$ and $\abseta < 2.4$, corresponding to a \enquote{Medium} \JVT WP.
 
The selection of jets containing $b$-hadrons~(\enquote{$b$-tagging}) is performed with a multivariate deep-learning algorithm~\cite{FTAG-2018-01,FTAG-2019-07}.
A selection that provides a 77\% efficiency for identifying jets containing $b$-hadrons~(\enquote{$b$-jets}) in simulated \ttbar events, with a rejection factor of \num{130} against
light-flavour jets and of five against jets containing $c$-hadrons, is used.
 
Associated scale factors are applied as multiplicative factors to the MC event weights, to correct for the mis-modelling of efficiencies associated with
the reconstruction, identification, isolation and trigger selection of electrons and muons, as well as the \JVT and $b$-tagging selection for jets.
 
The missing transverse momentum is defined as the negative vector sum of the transverse momenta of all selected and calibrated particle candidates~(electrons, muons and jets).
Low-momentum tracks from the primary vertex that are not associated with any of the reconstructed particle candidates described previously are also included as a \enquote{soft term}
in the calculation~\cite{PERF-2016-07}. The magnitude of the missing transverse momentum vector is denoted by \met.

Objects can satisfy both the jet and lepton selection criteria, and therefore a procedure called \enquote{overlap removal} is applied to avoid double counting the same energy clusters or tracks as different object types and to ensure that objects are matched to a unique hypothesis. If any electron candidate shares a track with a muon, the electron is removed. Similarly, if any jet is close to an electron (within $\DeltaR < 0.2$), the closest jet is removed. After that, electrons close to jets (within $\DeltaR < 0.4$) are removed to reduce the impact of non-prompt leptons. If the $\DeltaR$ between any remaining jet and a muon candidate is less than \num{0.4}, the muon candidate is removed if the jet has more than two associated tracks, otherwise, the jet is discarded. This sequential procedure is applied before the scale factors are determined for all objects, so any mis-modelling of the overlap removal is factored into the scale factors and their associated uncertainties.


\section{Event selection and definitions of control and signal regions}
\label{sec:regions}

Events were selected with either single-lepton or dilepton~(dielectron, dimuon and electron--muon) triggers with their minimum \pt thresholds varying from \num{12} to
\SI{26}{\GeV}, depending on the lepton flavour, the trigger type and the data-taking period~\cite{TRIG-2018-01,TRIG-2018-05}. A logical \enquote{OR} between the triggers
was applied. Only events with exactly three charged light leptons~(electrons or muons), as defined in Section~\ref{sec:objects}, are selected. If additional \emph{tight}
electrons or muons are found in the event, the event is rejected. The transverse momenta of the three charged leptons have to be larger than \num{30}, \num{20} and \SI{15}{\GeV}
for the leading, sub-leading and third lepton, respectively. A geometrical matching between these selected leptons and the
ones reconstructed by the trigger algorithms is required. Furthermore, the \pt of the lepton that fires a trigger needs to be above the \pt threshold of the respective trigger to ensure that the trigger
is maximally efficient.

The selected events are classified into four SRs, depending on their jet and $b$-jet multiplicities, as well as their \MET. In addition, four control regions (CRs) are defined to constrain the dominant backgrounds (\ttZ and \ttbar, \tW, or \Zjets events containing non prompt leptons from heavy flavour decays or \gam-conversions). In \ttW production with three charged leptons in the final state, two jets originating from $b$-quarks are expected from the hard process. Additional jet activity due to gluon radiation and showering effects can occur. However, background \ttZ events containing three leptons will typically contain extra jets from a hadronically-decaying top quark. Thus, the introduction of separate SRs with low and high jet multiplicities results in regions with improved signal over background ratios.
 
The definitions of the SRs and CRs are summarised in Table~\ref{tab:RegionsCuts}. The sum of the three lepton charges must be $\pm 1$. A requirement that the invariant mass
of the opposite-sign--same-flavour lepton pair~($m_{\ell\ell}^{\textrm{OSSF}}$) be at least \SI{30}{\GeV} is applied to remove the contributions from low-mass lepton
resonances~(e.g. $\Jpsi \to \ellell$). The number of \Zboson~boson candidates~(\NZ) is defined by the number of OSSF lepton pairs found in the event that have an invariant
mass in the range $[\mZ - \SI{10}{\GeV}, \mZ + \SI{10}{\GeV}]$. Exactly one \Zboson~boson candidate is required for the CR for the \ttZ background, but zero for all the other
regions. The CRs targeting non-prompt electron/muons arising from heavy-flavour~(HF) hadron decays~(\CRHFel and \CRHFmu) are separated by the flavour of the third lepton.
The third lepton must fail the isolation requirements in order to enrich these regions with HF leptons. The CR for \gam-conversions~(\CRConv) requires at least one of the
leptons to be an electron candidate that fails the $e/\gam$ ambiguity requirements defined in Section~\ref{sec:objects}. The contribution from electrons with misidentified
electric charge has been studied in MC and found to be negligible in the SRs.
 
Although the SR with two $b$-jets and low jet multiplicity~(\SRIIbL) is the most sensitive to \ttW production, the best result can be obtained from a statistical combination
of all SRs and CRs~(see Section~\ref{sec:results}).

\begin{table}[!htb]
\scriptsize
\def\arraystretch{1.4}
\caption{Summary of the requirements applied to define the signal and control regions of the analysis. \Nj includes $b$-tagged and non-$b$-tagged jets.
The labels \enquote{T} and \enquote{\noT} refer to \emph{tight} leptons that satisfy all selection requirements described in Section~\ref{sec:objects}~(T)
and \emph{loose} leptons that fail to satisfy the isolation requirements~(\noT).}
\label{tab:RegionsCuts}
\begin{center}
\begin{tabular}{c|c|c|c|c}
\toprule
& \multicolumn{4}{c}{Preselection} \\
\midrule
$\Nl \,(\ell=e/\mu)$ & \multicolumn{4}{c}{$=3$} \\
$\pt^{\ell}$~$(1^{\text{st}}/2^{\text{nd}}/3^{\text{rd}})$ & \multicolumn{4}{c}{$\geq \SI{30}{\GeV}$, $\geq \SI{20}{\GeV}$, $\geq \SI{15}{\GeV}$} \\
Sum of lepton charges & \multicolumn{4}{c}{$\pm 1$} \\
$m_{\ell\ell}^{\textrm{OSSF}}$ & \multicolumn{4}{c}{$\geq \SI{30}{\GeV}$} \\
\bottomrule
\toprule
& \multicolumn{4}{c}{Region-specific requirements} \\
\midrule
& \textbf{\SRIbL} & \textbf{\SRIbH} & \textbf{\SRIIbL} & \textbf{\SRIIbH} \\
\Nj & $[2,3]$ & $\geq4$ & $[2,3]$ & $\geq4$\\
\Nb & $=1$ & $=1$ & $\geq2$ & $\geq2$\\
\MET & $\geq \SI{50}{\GeV}$ & $\geq \SI{50}{\GeV}$ & -- & -- \\
\NZ & \multicolumn{4}{c}{$=0$}\\
Lepton criteria  & \multicolumn{4}{c}{TTT} \\
$e/\gam$ ambiguity-cuts & \multicolumn{4}{c}{satisfy all} \\
\midrule
& \textbf{\CRttZ} & \textbf{\CRHFel} & \textbf{\CRHFmu}  & \textbf{\CRConv}  \\
$\ell^{1\textrm{st}/2\textrm{nd}/3\textrm{rd}}$ & $\ell\ell\ell$   & $\ell\ell e$ & $\ell\ell\mu$ & $\ell \ell e$, $\ell e \ell$, $e \ell \ell$ \\
\Nj & $\geq4$ & $\geq2$ & $\geq2$ & $\geq2$  \\
\Nb & $\geq2$ & $=1$ & $=1$ & $\geq1$  \\
\MET & -- & $< \SI{50}{\GeV}$ & $< \SI{50}{\GeV}$ & $< \SI{50}{\GeV}$ \\
\NZ & $=1$ & $=0$ & $=0$ & $=0$ \\
Lepton criteria  & TTT & TT\noT & TT\noT & TTT \\
$e/\gam$ ambiguity-cuts & satisfy all & satisfy all & satisfy all & $\geq1$ fail \\
\bottomrule
\end{tabular}
\end{center}
\end{table}


\section{Lepton--top-quark matching}
\label{sec:leptontop}

In the \ttW process, the leptonic charge asymmetry is calculated only from the charged leptons that originate from the top quark and top antiquark. Since this search
targets events with three charged leptons, a problem arises when selecting the two leptons used to calculate the difference between their absolute pseudorapidities~(\dEta)
and ultimately the \lca value~(as defined in Eq.~(\ref{eq:lca_formula2})). The leptons that originate from a \ttbar pair always have opposite-sign charges. In a \ttW
event with three charged leptons, the two leptons with the same charge cannot both come from the \ttbar pair and will always contain one lepton from a top-quark or
top-antiquark decay and one lepton from the decay of the \Wboson boson produced in association with the \ttbar pair. This same-sign pair of leptons is henceforth referred
to as the \enquote{even} leptons. The remaining lepton with opposite charge, referred to as the \enquote{odd} lepton, will always originate from a top quark or top antiquark.
The problem of selecting the two leptons from top quark or top-antiquark decays is hence reduced to selecting one of the even leptons to calculate \dEta.
 
This problem is addressed using a boosted decision tree~(BDT) classifier algorithm that computes a discriminator value for each even lepton in each event. Large discriminator
values correspond to large probabilities that a given lepton originated from a top-quark decay. The lepton with the highest BDT discriminator score is selected to calculate
\dEta. The \dEta values calculated with the selected lepton and the odd lepton are denoted by \dEtaBDT. Five input variables that each discriminate between leptons from top-quark or
top-antiquark decays and leptons from associated \Wboson~boson decays are defined. They are the masses of the two systems formed from the lepton and the closest and second closest
$b$-jets, as well as the angular distances between the lepton and these $b$-jets: $m_{\ell b_{0}}$, $m_{\ell b_{1}}$, $\dR_{\ell b_{0}}$, $\dR_{\ell b_{1}}$. The last variable is
the lepton \pt. For events with only one $b$-tagged jet, if any of the remaining jets pass a looser $b$-tagging discriminant WP with respect to the default selection described in
Section~\ref{sec:objects}, the jet with the highest discriminant score is selected. If none of the other jets pass any $b$-tagging WP, the variables are constructed with the
closest untagged jet.
 
The training of the classifier is performed using the nominal \ttW \SHERPA MC sample. The BDT classifier is implemented using the \textsc{Scikit}-learn~\cite{scikit-learn}
package. A $k$-fold cross validation with five folds is used for the training and testing samples. The fraction of events in the \ttW sample in which the even lepton with the highest
BDT discriminator value originates from a top-quark or top-antiquark decay is estimated to be about 71\%, using the information from the MC event record.


\section{Systematic uncertainties in background and signal estimation}
\label{sec:systematics}

The predictions of the \ttW signal and the SM backgrounds are, in addition to the statistical uncertainties of their corresponding MC samples, affected by several sources of experimental
and theoretical systematic uncertainty. These uncertainties are classified into the different categories that are described in the following.
 
\subsection{Detector-related uncertainties}
 
Detector-related uncertainties include the simulation of pile-up events, the integrated luminosity, and effects related to the reconstruction and ID of the
particle candidates used in the analysis.
 
The uncertainty in the combined 2015--2018 integrated luminosity is 1.7\%~\cite{ATLAS-CONF-2019-021}, obtained using the LUCID-2 detector~\cite{Avoni:2018} for
the primary luminosity measurements. This systematic uncertainty affects all processes modelled using MC simulations apart from the processes where the
associated normalisation factors are obtained from data in the simultaneous fit~(see Section~\ref{sec:results}).
 
Uncertainties in the lepton reconstruction, ID, isolation and trigger efficiencies~\cite{EGAM-2018-01,PERF-2017-01,PERF-2015-10}, electron energy and muon momentum scale
and resolution~\cite{EGAM-2018-01,PERF-2015-10} are considered. Uncertainties associated with jets arise from the jet energy scale~(JES), the jet energy resolution~(JER)
and the \JVT jet selection requirement~\cite{JETM-2018-05}. In addition, uncertainties associated with the pile-up rejection~\cite{ATLAS-CONF-2014-018}, the scale and resolution
of the \MET~\cite{PERF-2016-07}, and the $b$-tagging efficiencies~\cite{ATL-PHYS-PUB-2017-013,FTAG-2018-01,ATLAS-CONF-2018-006} are considered.
 
\subsection{Signal and background modelling uncertainties}
\label{subsec:sys-theo}
 
Different sources of systematic uncertainty in the theoretical predictions of the \ttW signal are considered. To evaluate the effect of \muR and \muF uncertainties, the
scale factors used in the ME of the \SHERPA \ttW sample are varied simultaneously, as well as individually, by factors of \num{2.0} and \num{0.5} relative to their
nominal values~(but not including combinations where the variations differ by a factor of four). Uncertainties associated with the PDF sets are evaluated according
to the \PDFforLHC prescription~\cite{Butterworth:2015oua}. They include internal variations of the nominal PDF sets that are added in quadrature, uncertainties due
to the choice of PDF set, as well as variations of the \alphas parameter. The systematic uncertainties due to the modelling of the ME are evaluated by comparing the
prediction of the nominal MC sample with that of an alternative \ttW sample simulated with \MG~+~\PYTHIA and the \FXFX algorithm. Furthermore, to evaluate the systematic
uncertainties due to the PS, the hadronisation and the underlying event, the \ttW \POWHEG~+~\PYTHIA samples are compared with the \POWHEG~+~\HERWIG samples and their
relative differences are applied as uncertainties in the nominal \SHERPA prediction. As explained in Section~\ref{sec:samples}, the alternative samples employed for the
evaluation of modelling uncertainties are normalised to the same cross-sections as the respective nominal samples, so that the systematic uncertainties cover only
differences between the shapes of kinematic variable distributions, but not the overall normalisations of the processes.
 
For the theoretical systematic uncertainties in the \ttZ background, the same prescriptions as for the \ttW process are used to evaluate the effects of the \muR and
\muF uncertainties. For the systematic uncertainties due to the PS, the hadronisation and the underlying event, the nominal \ttZ prediction is compared with
that of an alternative \ttZ sample simulated with the same ME generator~(\MG), but interfaced to \HERWIG instead of \PYTHIA. Further alternative \ttZ samples
using a set of variations of the A14 tune's parameters are used to evaluate the uncertainty associated with the ISR, as mentioned in Section~\ref{sec:samples}.
 
For the \ttH, \tZq and \VVjets backgrounds, \muR and \muF uncertainties are considered. To evaluate the uncertainty in the ME, the PS, the hadronisation and the underlying
event of \ttH, the nominal prediction is compared with those from the alternative \ttH MC samples that use either a different ME generator~(\MG) or PS algorithm~(\HERWIG).
Furthermore, a normalisation uncertainty of $+5.8\%$ and $-9.2\%$ is applied for \ttH, following the NLO~(QCD + EW) calculations from Ref.~\cite{deFlorian:2016spz}. For the \tZq
process, a normalisation uncertainty of 14\% is applied, based on the dedicated ATLAS \tZq measurement described in Ref.~\cite{TOPQ-2018-01}. For the \VVjets backgrounds, a conservative
normalisation uncertainty of 20\% is used to account for differences in the quality of \VVjets modelling for different $b$-jet multiplicities. This uncertainty is derived from the level of agreement between data and MC simulation in several validation regions enriched in \VVjets and reflects the level of agreement between data and MC simulations in \cite{ATLAS:2019bsc}.
 
For the \ttbar and \Zjets backgrounds, which can only contribute to \lll final states via the presence of an additional fake or non-prompt lepton, \muR and \muF uncertainties
are also considered. For the \ttbar process, the PS, the hadronisation and the underlying event uncertainties are evaluated by comparing the nominal MC sample~(\POWHEG~+~\PYTHIA) with an
equivalent \ttbar sample with the same ME generator, but interfaced to \HERWIG. An extra uncertainty associated with the photon radiation in \ttbar events is applied by comparing the
predictions from \ttbar and \ttbar~+~\tty. The overlap between the photons radiated within the PS in the \ttbar~(\Zjets) simulation and the photons coming from \tty~(\Zy) is
removed from \ttbar~(\Zjets) for this comparison. This is done at particle level, where final-state photons that do not originate from prompt particle decays are removed if they are
inside the kinematic phase-space covered by the \tty~(\Zy) simulation. For \Zjets, the uncertainties in the ME, the PS, the hadronisation and the underlying event are evaluated from
a comparison between the nominal MC samples~(\POWHEG~+~\PYTHIA) and alternative \Zjets~(\Zy) samples, which are simulated with \SHERPA. The same approach as for \ttbar is followed to
account for the uncertainty associated with the photon radiation in \Zjets events.
 
For the other~(minor) background processes~(\tWZ, \ttWW, \HWZ, \VVV~($V=W/Z$), \ttt and \tttt), which typically contribute less than 2\% to the total event yields in the SRs,
a normalisation uncertainty of 30\% is applied. This is a conservative approach, based on the precision of the latest cross section measurement of \tttt \cite{ATLAS:2023ajo}, which should cover the known theoretical uncertainties of these backgrounds. The same also applies
to the MC sample generated with \ttW EW corrections at the order $\alpha^{3} \alphas$, as this is treated as a background~(see Section~\ref{sec:samples}).


\section{Extraction of the charge asymmetry at reconstruction level}
\label{sec:results}

To extract the leptonic charge asymmetry from the reconstructed leptons~(detector level), a simultaneous fit to the numbers of observed events in the SRs and CRs, as
defined in Section~\ref{sec:regions}, is performed. The fit is based on the profile-likelihood technique, with a binned likelihood function defined as a product of
Poisson probabilities of the observed event yields in all the regions. Systematic uncertainties~(see Section~\ref{sec:systematics}) are taken into account in the
likelihood function, each as a nuisance parameter constrained by a Gaussian probability density function~\cite{RooFit}.
 
The normalisation factors for the most relevant background processes in the SRs, namely \ttZ, non-prompt electrons/muons from HF decays and electrons from \gam-conversions,
are allowed to float freely in the fit. Events containing non-prompt leptons from HF decays and electrons from \gam-conversions are selected from processes that cannot contribute
directly to the \lll final state: \ttbar, \tW and \Zjets. These events are identified by requiring at least one lepton to originate from either a $b/c$-hadron~(HF$_{e/\mu}$) or a
converted photon~(\gam-conversion) according to the MC event records of the selected leptons. The variables used as input to the binned likelihood fit are the \pt of the third~(softest)
lepton in \CRHFel and \CRHFmu, as well as the scalar sum of the \pt of the selected jets in the event (\HT) in \CRttZ, since their distributions
show a sizeable shape difference between the targeted processes and the other SM backgrounds. In the SRs and \CRConv, the total numbers of events are used.
 
Each of the four SRs is separated into $\dEtaBDT \leq 0$~(\dEtaM) and $\dEtaBDT > 0$~(\dEtaP) regions. For the \dEtaM~(\dEtaP) set of regions, a single factor \NormDetaM~(\NormDetaP)
models the normalisations of the signal yields~(relatively to the SM cross-section) across the four SRs. Accordingly, the \lca value is extracted as a function of these normalisation
factors. Similarly, separate normalisation factors in the \dEtaM\ and \dEtaP\ sets of regions for the major background processes are allowed to float freely in the fit to avoid a bias
from an assumption of SM asymmetries for these processes in data.\footnote{The inclusive charge asymmetries at parton level for the simulated \ttZ and \ttbar samples are $\lca = -0.015$
and \num{0.004}, respectively.} An \enquote{injection test} is performed to verify that the fit result matches an injected non-SM \lca value and the fit can deal correctly with the fact
that the different SRs have different charge asymmetries.
 
\begin{figure}[!htpb]
\centering
\includegraphics[width=0.99\textwidth]{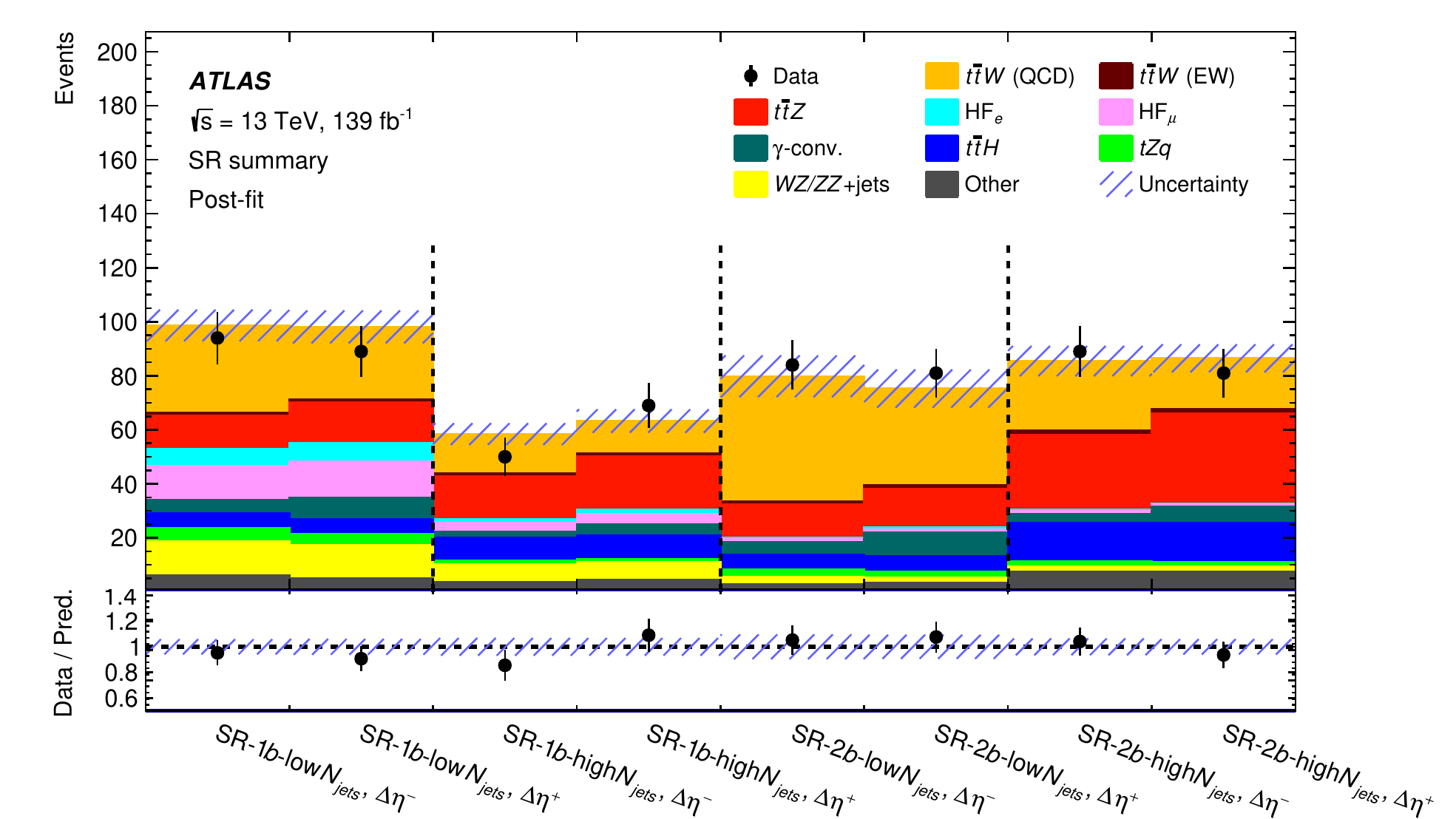}
\caption{Comparison between data and the post-fit predictions for $\dEtaBDT \leq 0$~(\dEtaM) and $\dEtaBDT > 0$~(\dEtaP) in the four SRs.
The error band includes the total uncertainties of the post-fit predictions, with the exception of the $\Delta\eta^{\pm}$ CR-dependency uncertainty. The ratio of the data to the total post-fit predictions is shown
in the lower panel.}
\label{fig:sr_postFit}
\end{figure}

\begin{figure}[!htpb]
\centering
\subfloat[]{\includegraphics[width=0.40\textwidth]{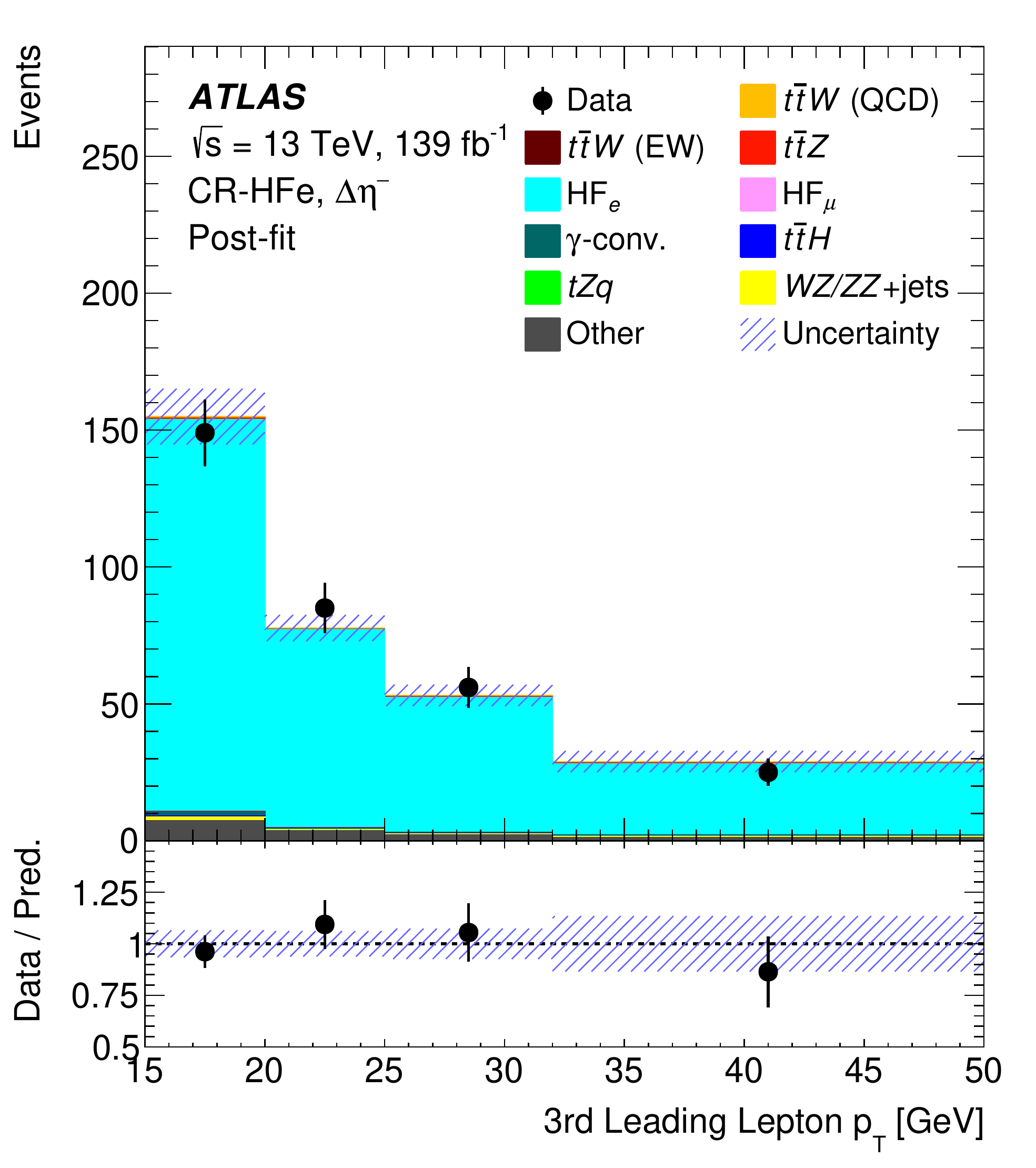}} \hspace{1cm}
\subfloat[]{\includegraphics[width=0.40\textwidth]{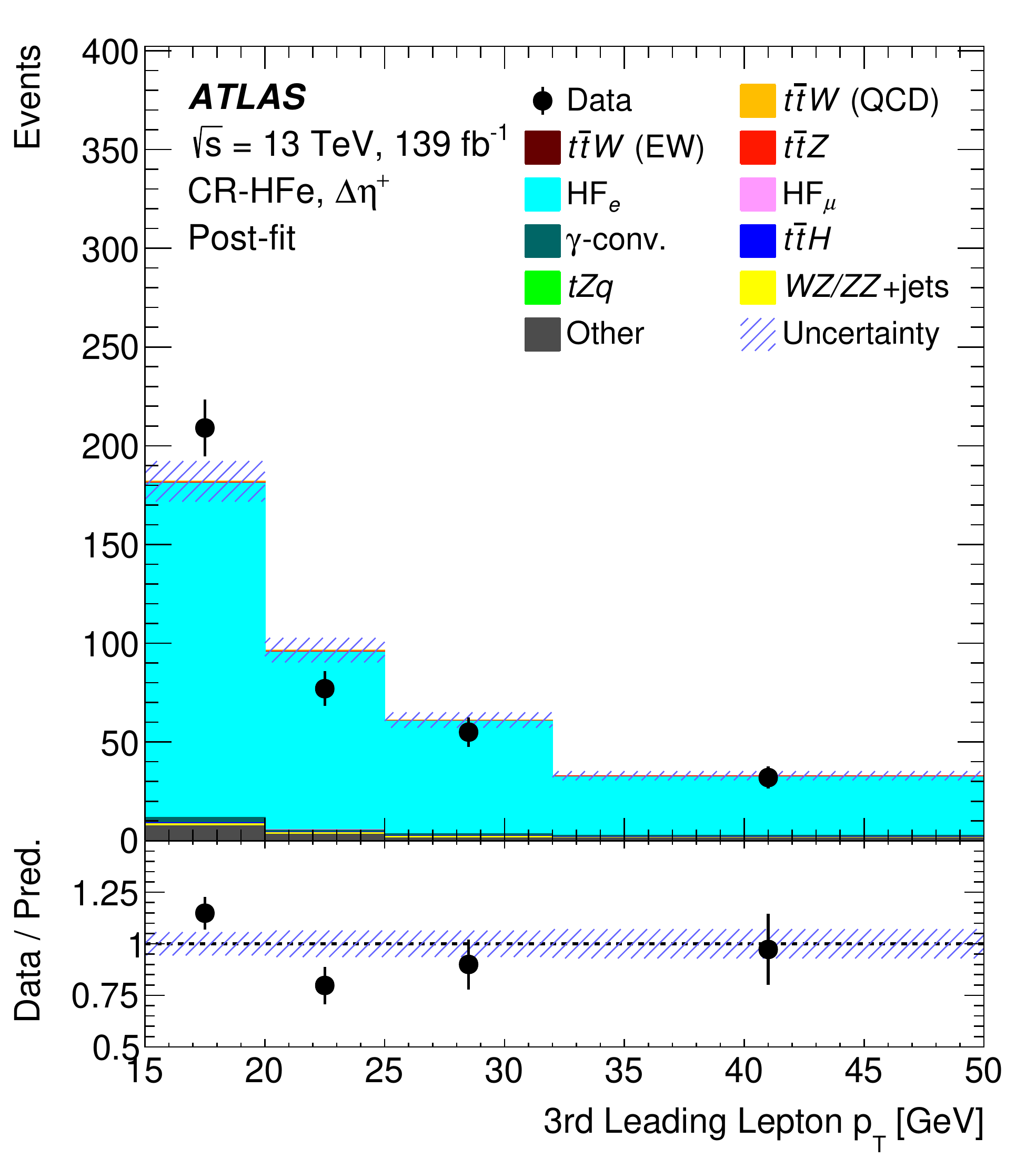}}
 
\subfloat[]{\includegraphics[width=0.40\textwidth]{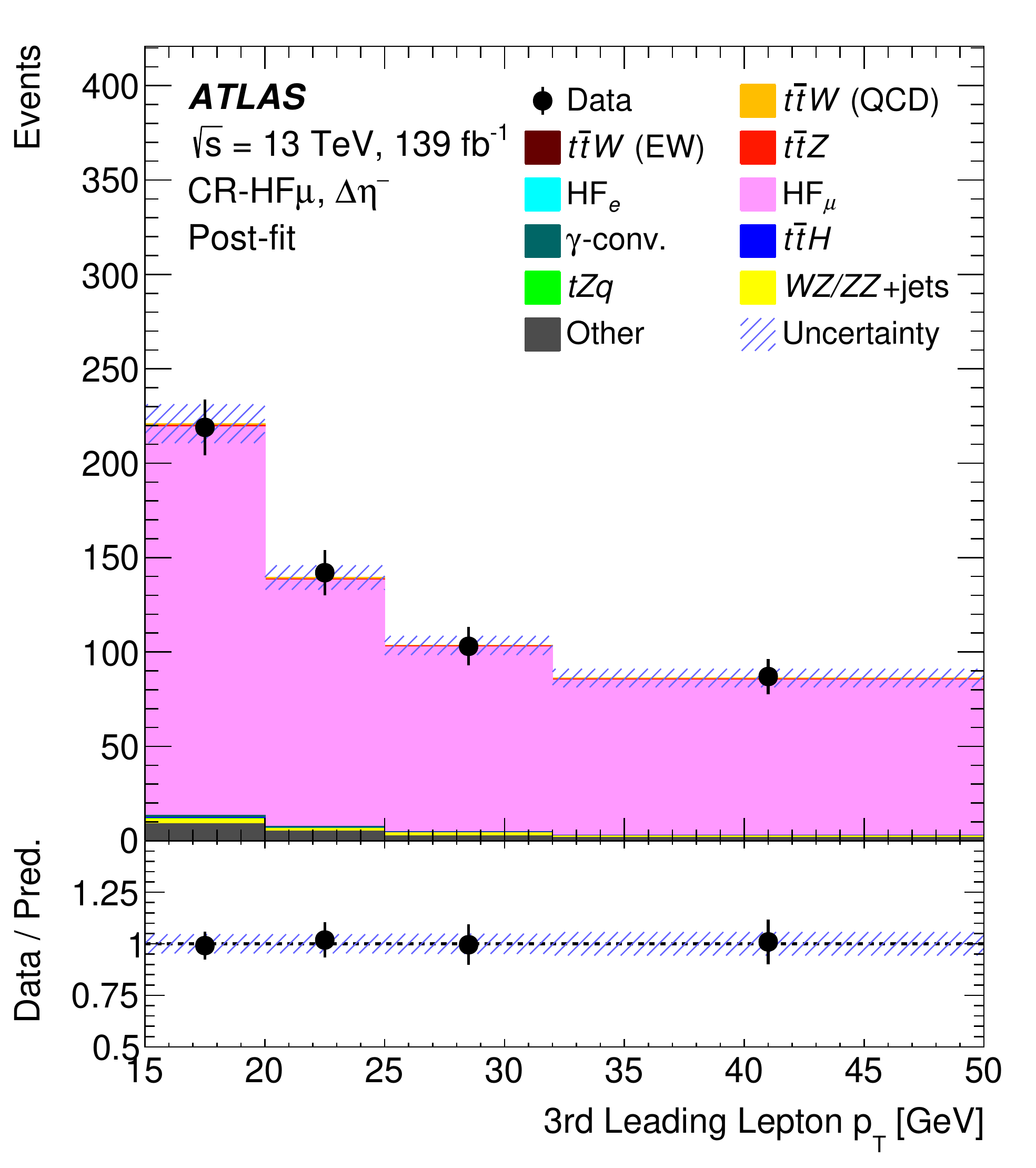}} \hspace{1cm}
\subfloat[]{\includegraphics[width=0.40\textwidth]{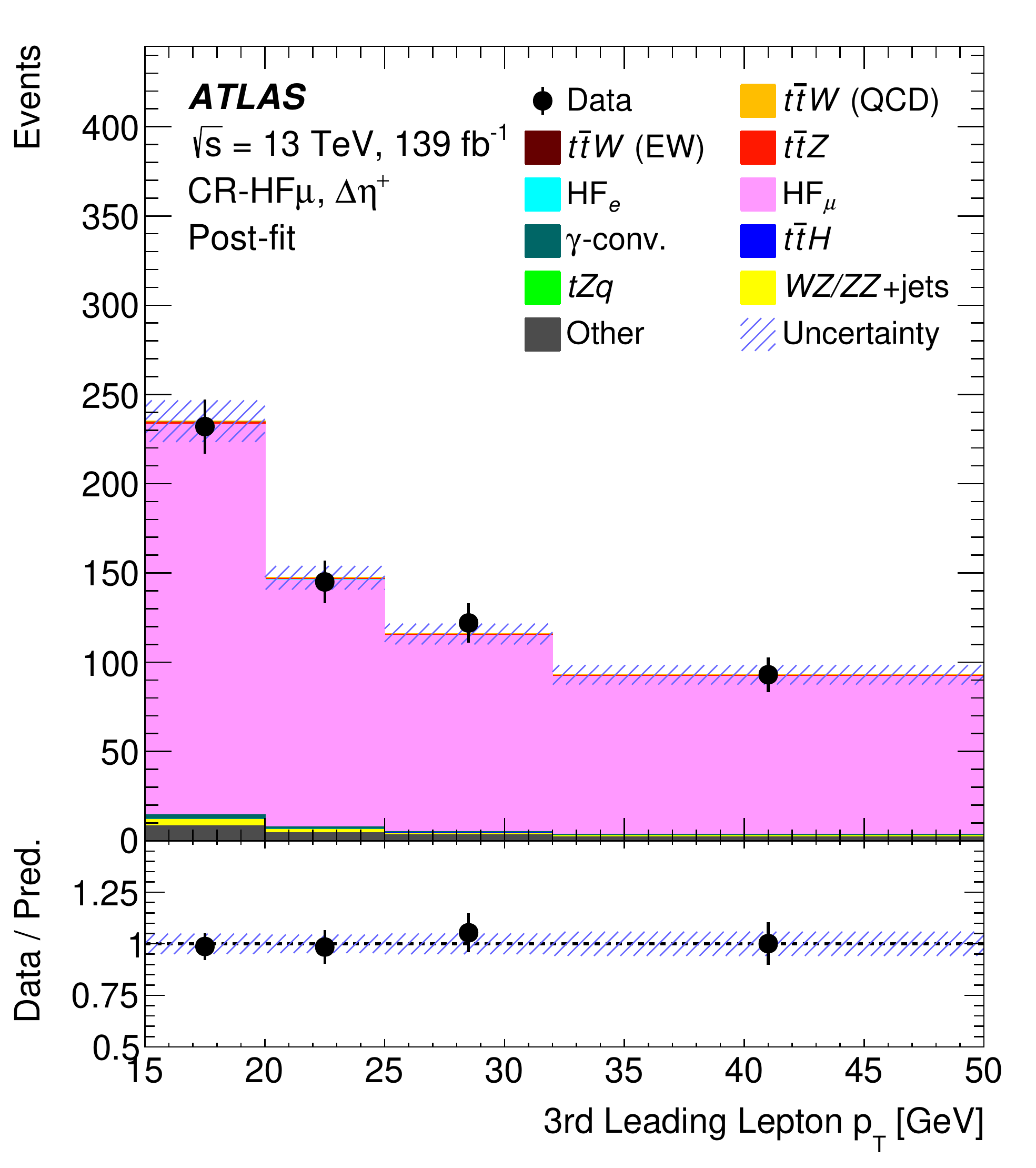}}
\caption{Comparison between data and the post-fit predictions in (a,b)~\CRHFel and (c,d)~\CRHFmu. The distributions show the \pt of the third lepton~(electron
or muon), which is the variable that is used for the binned likelihood fit. The regions are separated between $\dEtaBDT \leq 0$~(\dEtaM) and $\dEtaBDT > 0$~(\dEtaP).
The error bands include the total uncertainties in the post-fit predictions, with the exception of the $\Delta\eta^{\pm}$ CR-dependency uncertainty. The ratios of the data to the total post-fit predictions are shown in the lower
panels. Events with the \pt of the third lepton above \SI{50}{\GeV} are included in the rightmost bins.}
\label{fig:cr1_postFit}
\end{figure}

\begin{figure}[!htpb]
\centering
\subfloat[]{\includegraphics[width=0.40\textwidth]{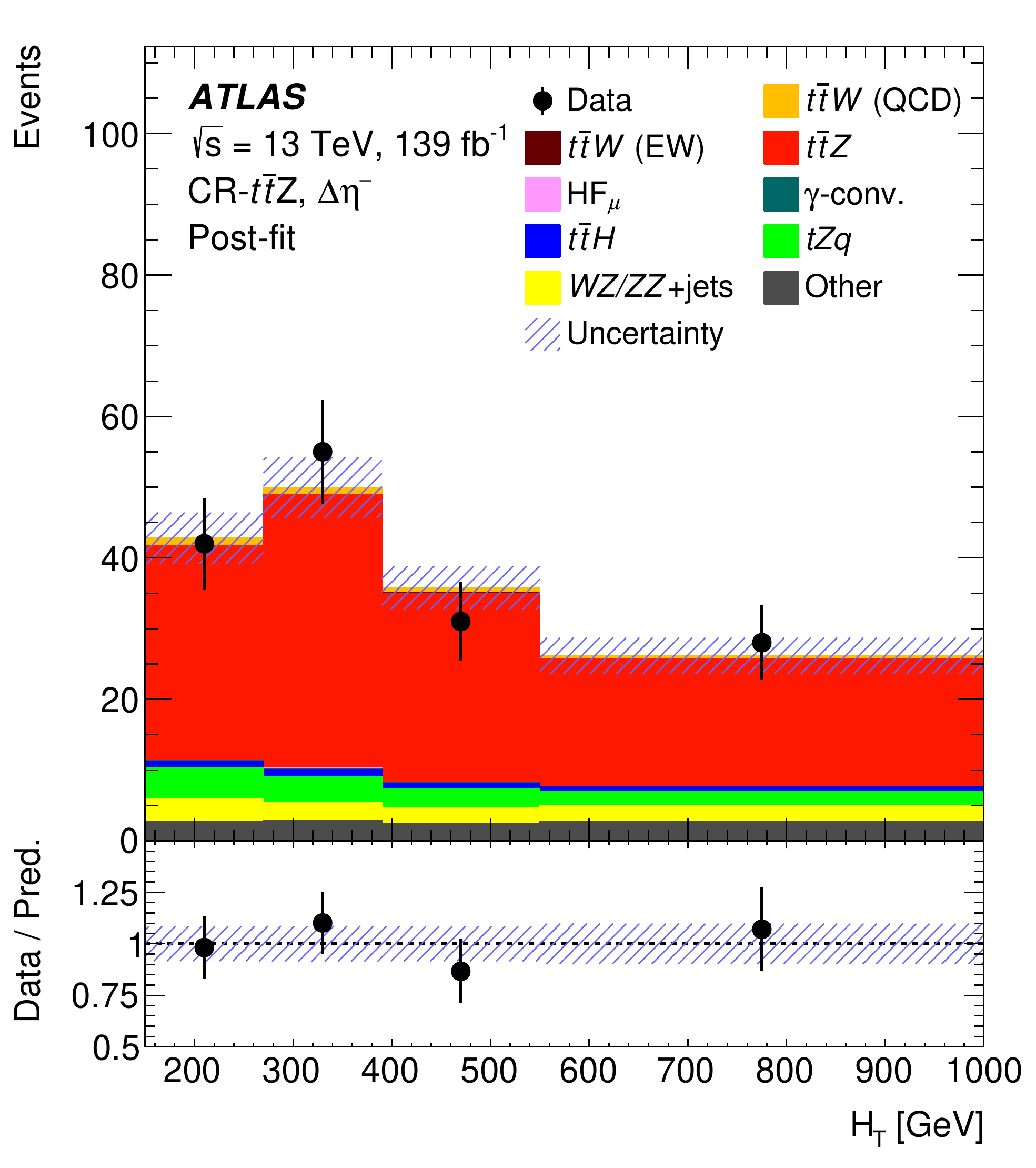}} \hspace{1cm}
\subfloat[]{\includegraphics[width=0.40\textwidth]{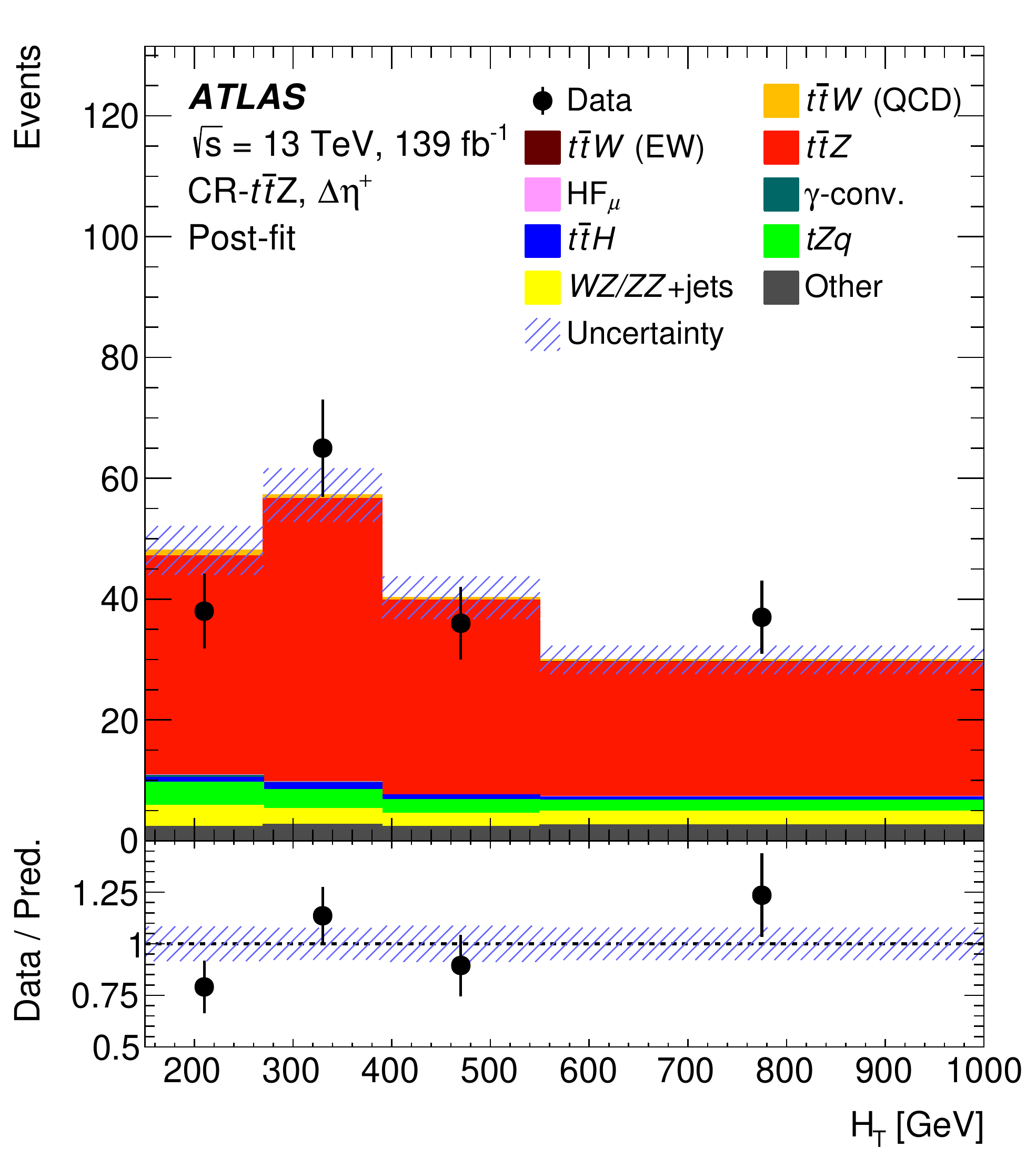}}
 
\subfloat[]{\includegraphics[width=0.60\textwidth]{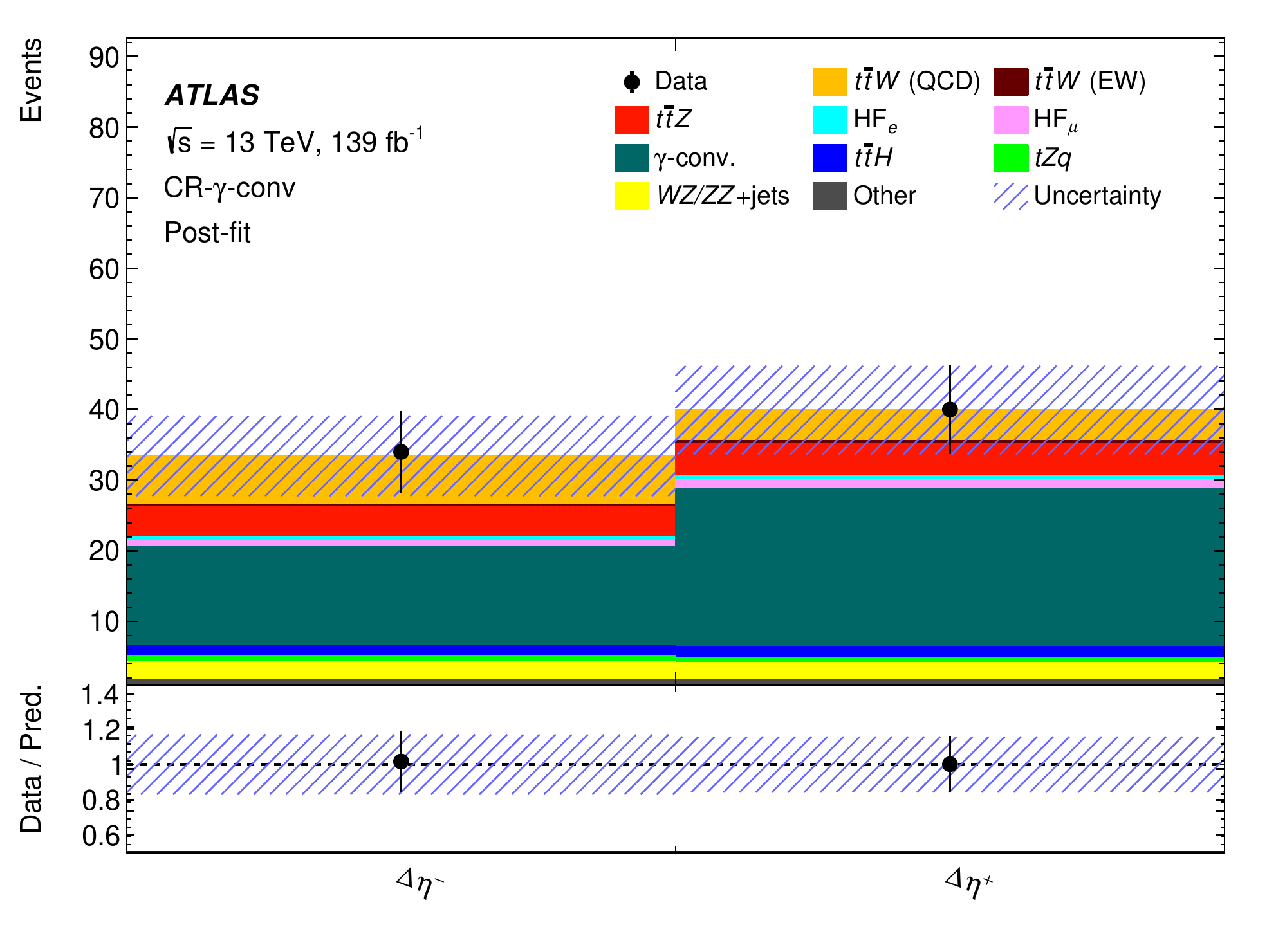}}
\caption{Comparison between data and the post-fit predictions in (a,b)~\CRttZ and (c)~\CRConv. The distributions are shown for the variables that are used
for the binned likelihood fit: \HT for \CRttZ and the total event yields for \CRConv. The regions are separated between $\dEtaBDT \leq 0$~(\dEtaM) and $\dEtaBDT > 0$~(\dEtaP).
The error bands include the total uncertainties in the post-fit predictions, with the exception of the $\Delta\eta^{\pm}$ CR-dependency uncertainty. The ratios of the data to the total post-fit predictions are shown in the lower
panels. Events with an \HT above \SI{1}{\TeV} are included in the rightmost bins of (a)~and~(b).}
\label{fig:cr2_postFit}
\end{figure}

The predicted and observed numbers of events in the SRs and CRs before performing the simultaneous fit~(\enquote{pre-fit}) are shown in Table~\ref{tab:yields_preFit}.
The indicated uncertainties consider statistical as well as all experimental and theoretical systematic uncertainties described in Section~\ref{sec:systematics}.
The numbers of events in the SRs and CRs after the fit to data~(\enquote{post-fit}) are given in Table~\ref{tab:yields_postFit}. Comparisons between data and the post-fit predictions for \dEtaM\ and \dEtaP\ in the four SRs are shown in Figure~\ref{fig:sr_postFit}. Data and the
post-fit SM predictions for the variables that are used for the binned likelihood fit are depicted in Figure~\ref{fig:cr1_postFit} for \CRHFel and \CRHFmu, and in Figure~\ref{fig:cr2_postFit} for \CRttZ and \CRConv.
 
\begin{table}[!htb]
\def\arraystretch{1.3}
\caption{The predicted and observed numbers of events in the control and signal regions. The predictions are shown before the fit to data. The indicated
uncertainties consider statistical as well as all experimental and theoretical systematic uncertainties, with the exception of the $\Delta\eta^{\pm}$ CR-dependency uncertainty. Background categories with event yields shown as \enquote{–-}
do contribute less than \num{0.01} to a region.}
\label{tab:yields_preFit}
\begin{center}
\resizebox{\textwidth}{!}{
\begin{tabular}{l | S[table-format=3.2]@{\(\,\pm\,\)}S[table-format=2.2] S[table-format=3.2]@{\(\,\pm\,\)}S[table-format=2.2] | S[table-format=3.2]@{\(\,\pm\,\)}S[table-format=2.2] S[table-format=3.2]@{\(\,\pm\,\)}S[table-format=2.2] | S[table-format=3.2]@{\(\,\pm\,\)}S[table-format=2.2] S[table-format=3.2]@{\(\,\pm\,\)}S[table-format=2.2] | S[table-format=3.2]@{\(\,\pm\,\)}S[table-format=2.2] S[table-format=3.2]@{\(\,\pm\,\)}S[table-format=2.2]}
\toprule
\multirow{2}{*}{{Process}} & \multicolumn{4}{c|}{{\CRttZ}} & \multicolumn{4}{c|}{{\CRHFel}} & \multicolumn{4}{c|}{{\CRHFmu}} & \multicolumn{4}{c}{{\CRConv}} \\
& \multicolumn{2}{c}{{\dEtaM}} & \multicolumn{2}{c|}{{\dEtaP}} & \multicolumn{2}{c}{{\dEtaM}} & \multicolumn{2}{c|}{{\dEtaP}} & \multicolumn{2}{c}{{\dEtaM}} & \multicolumn{2}{c|}{{\dEtaP}} & \multicolumn{2}{c}{{\dEtaM}} & \multicolumn{2}{c}{{\dEtaP}} \\
\midrule
\ttW~(QCD) & 1.8 & 0.4 & 1.49 & 0.19 & 1.18 & 0.19 & 1.13 & 0.18 & 1.72 & 0.20 & 1.37 & 0.28 & 4.1 & 0.7 & 2.92 & 0.18 \\
\ttW~(EW)  & 0.18 & 0.07 & 0.16 & 0.06 & 0.10 & 0.04 & 0.09 & 0.04 & 0.09 & 0.04 & 0.14 & 0.05 & 0.23 & 0.08 & 0.36 & 0.12 \\
\ttZ       & 107 & 6 & 107 & 6 & 1.42 & 0.23 & 1.5 & 0.4 & 2.20 & 0.23 & 2.00 & 0.14 & 4.04 & 0.19 & 3.65 & 0.32 \\
HF$_{e}$   & \multicolumn{2}{c}{{--}}  & \multicolumn{2}{c|}{{--}} & 350 & 40 & 362 & 27 & 0.18 & 0.11 & 0.20 & 0.09 & 1.0 & 0.6 & 0.67 & 0.35 \\
HF$_{\mu}$ & 0.14 & 0.08 & 0.19 & 0.09 & 0.20 & 0.09 & 0.28 & 0.10 & 520 & 40 & 530 & 50 & 0.9 & 0.5 & 1.1 & 0.9 \\
\gam-conv. & 0.55 & 0.14 & 0.41 & 0.13 & 3.8 & 2.5 & 4.7 & 2.9 & 2.6 & 2.4 & 3.3 & 2.5 & 18.8 & 1.4 & 17.5 & 1.3 \\
\ttH       & 3.3 & 0.4 & 3.20 & 0.32 & 0.87 & 0.13 & 0.89 & 0.11 & 1.18 & 0.11 & 1.22 & 0.22 & 1.48 & 0.20 & 1.5 & 0.4 \\
\tZq       & 12.6 & 2.2 & 11.0 & 1.9 & 0.48 & 0.11 & 0.43 & 0.09 & 0.95 & 0.18 & 0.81 & 0.15 & 0.68 & 0.12 & 0.70 & 0.13 \\
\VVjets    & 12 & 4 & 12 & 4 & 3.0 & 0.9 & 3.3 & 1.0 & 7.2 & 2.4 & 7.9 & 2.5 & 3.1 & 0.9 & 2.9 & 0.8 \\
Other      & 10.7 & 3.3 & 10.2 & 3.3 & 14 & 4 & 13 & 5 & 17 & 7 & 17 & 6 & 1.6 & 0.8 & 1.5 & 0.6 \\
\midrule
SM total   & 148 & 10 & 146 & 10 & 380 & 40 & 387 & 28 & 550 & 40 & 560 & 50 & 35.9 & 2.4 & 32.9 & 2.3 \\
Data       & \multicolumn{2}{c}{156} & \multicolumn{2}{c|}{176} & \multicolumn{2}{c}{315} & \multicolumn{2}{c|}{373} & \multicolumn{2}{c}{551} & \multicolumn{2}{c|}{592} & \multicolumn{2}{c}{34} & \multicolumn{2}{c}{40} \\
\bottomrule \multicolumn{16}{c}{} \\ \toprule
\multirow{2}{*}{{Process}} & \multicolumn{4}{c|}{{\SRIbL}} & \multicolumn{4}{c|}{{\SRIbH}} & \multicolumn{4}{c|}{{\SRIIbL}} & \multicolumn{4}{c}{{\SRIIbH}} \\
& \multicolumn{2}{c}{{\dEtaM}} & \multicolumn{2}{c|}{{\dEtaP}} & \multicolumn{2}{c}{{\dEtaM}} & \multicolumn{2}{c|}{{\dEtaP}} & \multicolumn{2}{c}{{\dEtaM}} & \multicolumn{2}{c|}{{\dEtaP}} & \multicolumn{2}{c}{{\dEtaM}} & \multicolumn{2}{c}{{\dEtaP}} \\
\midrule
\ttW~(QCD) & 19.0 & 2.8 & 17 & 4 & 9.2 & 1.1 & 8.2 & 1.1 & 25 & 7 & 21 & 6 & 14.7 & 3.4 & 12.2 & 1.9 \\
\ttW~(EW)  & 1.06 & 0.34 & 1.3 & 0.4 & 1.05 & 0.34 & 1.07 & 0.34 & 1.2 & 0.4 & 1.3 & 0.4 & 1.8 & 0.6 & 1.6 & 0.5 \\
\ttZ       & 12.0 & 1.0 & 12.1 & 1.1 & 15.5 & 1.4 & 15.5 & 1.1 & 11.4 & 1.4 & 10.8 & 1.4 & 26.2 & 1.8 & 25.8 & 1.7 \\
HF$_{e}$   & 7.2 & 1.2 & 7.5 & 1.5 & 1.7 & 0.7 & 1.6 & 0.6 & 0.7 & 0.5 & 0.6 & 0.5 & 0.69 & 0.35 & 0.37 & 0.19 \\
HF$_{\mu}$ & 12.5 & 2.0 & 13 & 4 & 3.2 & 0.8 & 3.5 & 1.3 & 1.35 & 0.34 & 1.11 & 0.33 & 1.0 & 0.4 & 0.9 & 0.5 \\
\gam-conv. & 6.7 & 0.9 & 6.1 & 1.0 & 3.1 & 0.5 & 3.4 & 0.8 & 6.1 & 0.8 & 6.9 & 0.8 & 4.4 & 0.7 & 4.6 & 0.6 \\
\ttH       & 5.5 & 0.8 & 5.6 & 0.8 & 8.6 & 0.8 & 8.7 & 0.9 & 5.5 & 1.1 & 5.5 & 1.0 & 14.1 & 1.8 & 14.2 & 1.7 \\
\tZq       & 5.1 & 0.9 & 4.2 & 0.7 & 1.40 & 0.31 & 1.15 & 0.27 & 2.8 & 0.5 & 2.3 & 0.4 & 1.92 & 0.34 & 1.64 & 0.30 \\
\VVjets    & 15 & 4 & 14 & 4 & 8.0 & 2.8 & 7.6 & 2.5 & 2.9 & 0.9 & 2.2 & 0.7 & 2.2 & 0.7 & 2.2 & 0.7 \\
Other      &  5.6 & 2.0 & 5.1 & 1.6 & 4.5 & 2.4 & 4.7 & 1.5 & 2.6 & 1.1 & 2.9 & 1.3 & 10 & 6 & 9 & 5 \\
\midrule
SM total   & 89 & 6 & 85 & 7 & 56 & 6 & 56 & 6 & 59 & 9 & 55 & 7 & 77 & 8 & 73 & 7 \\
Data       & \multicolumn{2}{c}{94} & \multicolumn{2}{c|}{89} & \multicolumn{2}{c}{50} & \multicolumn{2}{c|}{69} & \multicolumn{2}{c}{84} & \multicolumn{2}{c|}{81} & \multicolumn{2}{c}{89} & \multicolumn{2}{c}{81} \\
\bottomrule
\end{tabular} }
\end{center}
\end{table}
 
\begin{table}[!htb]
\def\arraystretch{1.3}
\caption{The predicted and observed numbers of events in the control and signal regions. The predictions are shown after the fit to data. The indicated
uncertainties consider statistical as well as all experimental and theoretical systematic uncertainties, with the exception of the $\Delta\eta^{\pm}$ CR-dependency uncertainty. Background categories with event yields shown as \enquote{–-}
do contribute less than \num{0.01} to a region.}
\label{tab:yields_postFit}
\begin{center}
\resizebox{\textwidth}{!}{
\begin{tabular}{l | S[table-format=3.3]@{\(\,\pm\,\)}S[table-format=2.3] S[table-format=3.3]@{\(\,\pm\,\)}S[table-format=2.3] | S[table-format=3.3]@{\(\,\pm\,\)}S[table-format=2.3] S[table-format=3.3]@{\(\,\pm\,\)}S[table-format=2.3] | S[table-format=3.2]@{\(\,\pm\,\)}S[table-format=2.2] S[table-format=3.2]@{\(\,\pm\,\)}S[table-format=2.2] | S[table-format=3.2]@{\(\,\pm\,\)}S[table-format=2.2] S[table-format=3.2]@{\(\,\pm\,\)}S[table-format=2.2]}
\toprule
\multirow{2}{*}{{Process}} & \multicolumn{4}{c|}{{\CRttZ}} & \multicolumn{4}{c|}{{\CRHFel}} & \multicolumn{4}{c|}{{\CRHFmu}} & \multicolumn{4}{c}{{\CRConv}} \\
& \multicolumn{2}{c}{{\dEtaM}} & \multicolumn{2}{c|}{{\dEtaP}} & \multicolumn{2}{c}{{\dEtaM}} & \multicolumn{2}{c|}{{\dEtaP}} & \multicolumn{2}{c}{{\dEtaM}} & \multicolumn{2}{c|}{{\dEtaP}} & \multicolumn{2}{c}{{\dEtaM}} & \multicolumn{2}{c}{{\dEtaP}} \\
\midrule
\ttW~(QCD) & 3.2 & 0.7 & 2.2 & 0.7 & 1.8 & 0.5 & 1.7 & 0.5 & 2.6 & 0.8 & 1.8 & 0.8 & 7.0 & 1.3 & 4.4 & 1.3 \\
\ttW~(EW)  & 0.18 & 0.06 & 0.16 & 0.05 & 0.10 & 0.03 & 0.09 & 0.03 & 0.09 & 0.03 & 0.14 & 0.04 & 0.23 & 0.07 & 0.36 & 0.11 \\
\ttZ       & 114 & 13 & 138 & 14 & 1.45 & 0.27 & 1.7 & 0.4 & 2.3 & 0.4 & 2.55 & 0.35 & 4.3 & 0.6 & 4.6 & 0.6 \\
HF$_{e}$   & \multicolumn{2}{c}{{--}} & \multicolumn{2}{c|}{{--}} & 290 & 18 & 346 & 20 & 0.15 & 0.02 & 0.19 & 0.02 & 0.59 & 0.27 & 0.52 & 0.17 \\
HF$_{\mu}$ & 0.133 & 0.012 & 0.201 & 0.020 & 0.195 & 0.018 & 0.277 & 0.029 & 516 & 25 & 556 & 25 & 0.8 & 0.4 & 1.3 & 0.8 \\
\gam-conv. & 0.40 & 0.18 & 0.52 & 0.16 & 2.8 & 2.2 & 6 & 4 & 1.9 & 2.0 & 4.2 & 3.4 & 14 & 6 & 22 & 7 \\
\ttH       & 3.3 & 0.4 & 3.23 & 0.31 & 0.86 & 0.13 & 0.87 & 0.10 & 1.16 & 0.11 & 1.19 & 0.22 & 1.49 & 0.20 & 1.6 & 0.4 \\
\tZq       & 12.6 & 2.2 & 11.0 & 1.9 & 0.47 & 0.10 & 0.42 & 0.08 & 0.95 & 0.17 & 0.79 & 0.14 & 0.68 & 0.11 & 0.70 & 0.12 \\
\VVjets    & 10.2 & 2.9 & 10.6 & 3.1 & 2.6 & 0.7 & 2.8 & 0.7 & 6.3 & 1.7 & 6.7 & 1.8 & 2.6 & 0.7 & 2.5 & 0.6 \\
Other      & 10.8 & 3.2 & 10.0 & 2.9 & 14 & 4 & 13 & 5 & 18 & 7 & 18 & 6 & 1.7 & 0.8 & 1.7 & 0.6 \\
\midrule
SM total   & 155 & 12 & 175 & 13 & 315 & 18 & 373 & 19 & 550 & 23 & 591 & 24 & 33 & 6 & 40 & 6 \\
Data       & \multicolumn{2}{c}{156} & \multicolumn{2}{c|}{176} & \multicolumn{2}{c}{315} & \multicolumn{2}{c|}{373} & \multicolumn{2}{c}{551} & \multicolumn{2}{c|}{592} & \multicolumn{2}{c}{34} & \multicolumn{2}{c}{40} \\
\bottomrule \multicolumn{16}{c}{} \\ \toprule
\multirow{2}{*}{{Process}} & \multicolumn{4}{c|}{{\SRIbL}} & \multicolumn{4}{c|}{{\SRIbH}} & \multicolumn{4}{c|}{{\SRIIbL}} & \multicolumn{4}{c}{{\SRIIbH}} \\
& \multicolumn{2}{c}{{\dEtaM}} & \multicolumn{2}{c|}{{\dEtaP}} & \multicolumn{2}{c}{{\dEtaM}} & \multicolumn{2}{c|}{{\dEtaP}} & \multicolumn{2}{c}{{\dEtaM}} & \multicolumn{2}{c|}{{\dEtaP}} & \multicolumn{2}{c}{{\dEtaM}} & \multicolumn{2}{c}{{\dEtaP}} \\
\midrule
 
\ttW~(QCD) & 32 & 6 & 27 & 6 & 14 & 4 & 12.1 & 3.4 & 46 & 9 & 36 & 8 & 26 & 6 & 19 & 5 \\
\ttW~(EW)  & 1.04 & 0.32 & 1.3 & 0.4 & 1.04 & 0.32 & 1.05 & 0.32 & 1.2 & 0.4 & 1.3 & 0.4 & 1.8 & 0.5 & 1.6 & 0.5 \\
\ttZ       & 12.4 & 2.0 & 15.0 & 2.2 & 16.0 & 2.2 & 19.6 & 2.3 & 12.3 & 2.3 & 14.3 & 2.6 & 27.6 & 3.3 & 33.2 & 3.5 \\
HF$_{e}$   & 6.4 & 1.0 & 6.8 & 0.8 & 1.5 & 0.5 & 1.7 & 0.4 & 0.40 & 0.20 & 0.79 & 0.35 & 0.45 & 0.14 & 0.39 & 0.14 \\
HF$_{\mu}$ & 12.5 & 1.5 & 13.6 & 2.5 & 3.1 & 0.6 & 3.6 & 0.9 & 1.30 & 0.23 & 1.19 & 0.19 & 1.04 & 0.29 & 0.9 & 0.5 \\
\gam-conv. & 4.9 & 2.3 & 7.7 & 2.6 & 2.3 & 1.1 & 4.3 & 1.6 & 4.6 & 2.1 & 8.8 & 2.9 & 3.3 & 1.5 & 5.9 & 1.9 \\
\ttH       & 5.4 & 0.8 & 5.5 & 0.8 & 8.4 & 0.8 & 8.6 & 0.8 & 5.5 & 1.1 & 5.6 & 1.0 & 14.3 & 1.7 & 14.4 & 1.7 \\
\tZq       & 5.0 & 0.9 & 4.1 & 0.7 & 1.38 & 0.27 & 1.16 & 0.24 & 2.8 & 0.5 & 2.3 & 0.4 & 1.93 & 0.33 & 1.65 & 0.29 \\
\VVjets    & 12.6 & 3.0 & 12.3 & 3.0 & 6.7 & 2.0 & 6.5 & 1.8 & 2.5 & 0.7 & 1.9 & 0.5 & 1.9 & 0.6 & 1.9 & 0.5 \\
Other      & 6.0 & 2.1 & 5.2 & 1.6 & 3.6 & 1.8 & 4.6 & 1.4 & 2.9 & 1.2 & 3.3 & 1.3 & 8 & 4 & 8 & 4 \\
\midrule
SM total   & 99 & 6 & 98 & 6 & 58 & 4 & 63 & 4 & 80 & 8 & 75 & 7 & 85 & 6 & 86 & 5 \\
Data       & \multicolumn{2}{c}{94} & \multicolumn{2}{c|}{89} & \multicolumn{2}{c}{50} & \multicolumn{2}{c|}{69} & \multicolumn{2}{c}{84} & \multicolumn{2}{c|}{81} & \multicolumn{2}{c}{89} & \multicolumn{2}{c}{81} \\
\bottomrule
\end{tabular} }
\end{center}
\end{table}
 
\begin{figure}[!htpb]
\centering
\includegraphics[width=0.40\textwidth]{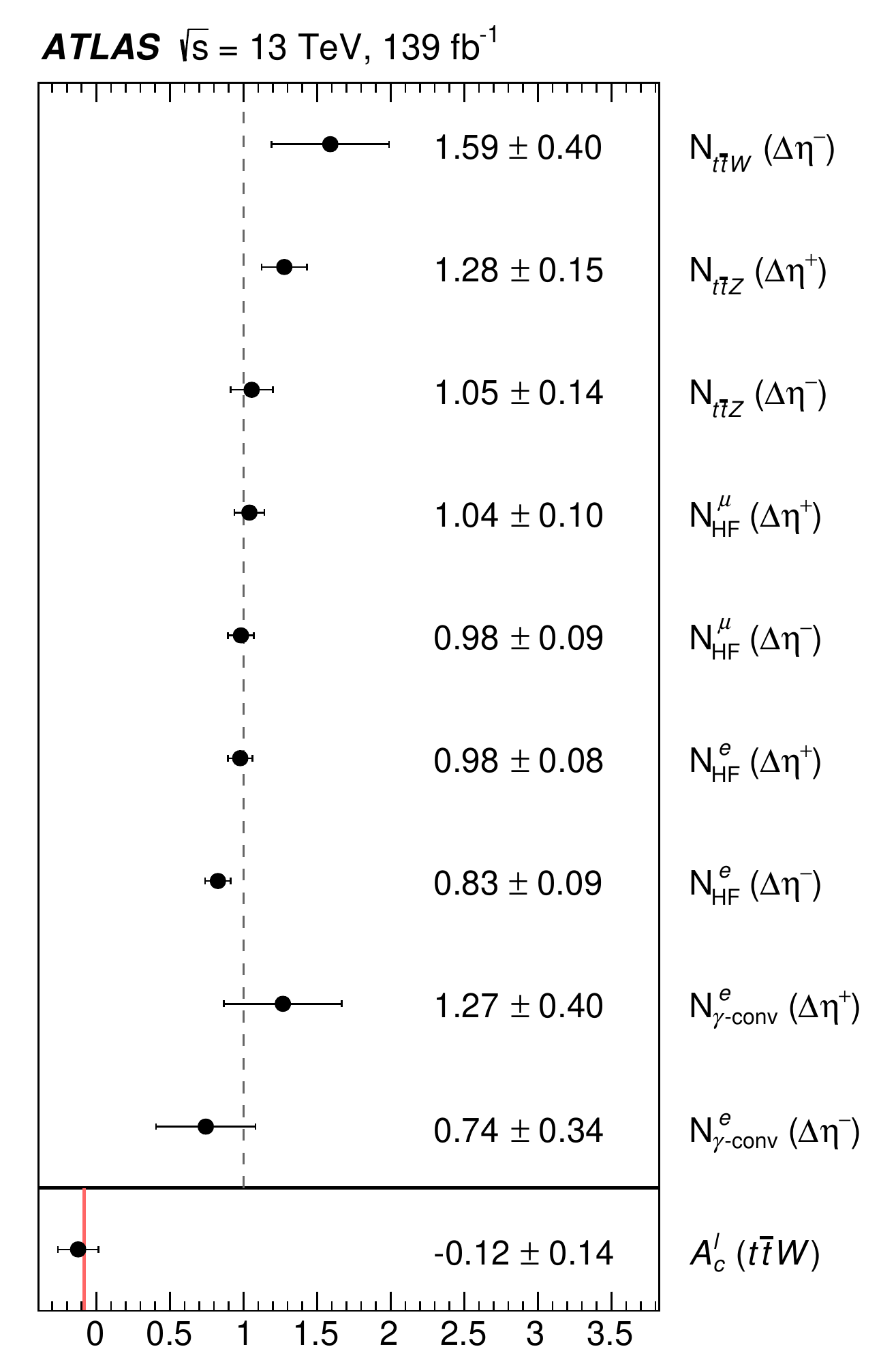}
\caption{Normalisation factors for the major background processes, together with \NormDetaM for \ttW and the \lca value extracted from the fit to data in the CRs and SRs at detector level.
The normalisation factors, \NormttZ, \NormCO, \NormHFe and \NormHFmu, are obtained separately for $\dEtaBDT \leq 0$~(\dEtaM) and $\dEtaBDT > 0$~(\dEtaP). The indicated
uncertainties consider statistical as well as systematic uncertainties, with the exception of the $\Delta\eta^{\pm}$ CR-dependency uncertainty. The solid vertical line in the last entry shows the \lca SM expectation, calculated using the
\ttW \SHERPA simulation.}
\label{fig:normFactors_reco}
\end{figure}

The normalisation factors for the major background processes, \NormttZ, \NormCO, \NormHFe and \NormHFmu~(all obtained separately for \dEtaM\ and \dEtaP), together with \NormDetaM
and the \lca value for the \ttW signal, are given in Figure~\ref{fig:normFactors_reco}. The normalisation factor for the \ttW process was checked and found to be~(within its
uncertainty) compatible with the latest ATLAS and CMS \ttW cross-section measurements~\cite{TOPQ-2016-11,CMS:2022tkv}. Tests using MC simulation were also performed to validate
that the extracted \lca value is not biased by the absolute normalisation of the \ttW process.
 
The normalisation factors for some of the background processes~(in particular \NormttZ and \NormCO) show small differences between \dEtaM\ and \dEtaP. As these processes are not expected
to have significant charge asymmetries in the SM at this level of precision, there is an uncertainty on how best to model this data. In the nominal fit, due to the independent normalisation
factors for \dEtaM\ and \dEtaP\ in the CRs, the observed background asymmetries are precisely modelled. To account for the possibility that the observed asymmetry is due to a statistical fluctuation,
an alternative fit is performed where only one normalisation factor is assigned to each of these processes~(thus fixing their asymmetries to the SM expectation). The difference between the
results of these two fit set-ups is assigned as an extra systematic uncertainty in the extracted \lca value. This uncertainty~(denoted as $\Delta\eta^{\pm}$ CR-dependency) is found to
be \num{0.05} and is one of the leading systematic uncertainties.

The leptonic charge asymmetry in \ttW is found to be
\begin{equation*}
\lca(\ttW) = -0.12 \pm 0.14 \,\text{(stat.)} \pm 0.05 \,\text{(syst.)}.
\end{equation*}
This is consistent with the SM expectation of
\begin{equation*}
\lca(\ttW)_{\text{SM}} = -0.084 \, ^{+0.005}_{-0.003} \,\text{(scale)} \pm 0.006 \,\text{(MC stat.)},
\end{equation*}
calculated using the nominal \ttW \SHERPA simulation. The contributions from the most relevant uncertainties are summarised in Table~\ref{tab:syst_reco}. The uncertainties are symmetrised
and grouped into several type-related categories and are shown together with the total systematic and statistical uncertainties. The dominant systematic uncertainties are the
$\Delta\eta^{\pm}$ CR-dependency, the JER, as well as the modelling uncertainties of the \ttW and \ttZ MC processes detailed in Section~\ref{sec:systematics}. Overall, the result is
limited by the statistical uncertainty of the data.
 
\FloatBarrier

\begin{table}[!htb]
\scriptsize
\def\arraystretch{1.2}
\caption{List of the most relevant systematic and statistical uncertainties in the extracted leptonic charge asymmetry $\lca(\ttW)$ from the simultaneous fit.
For this table, the uncertainties are symmetrised and grouped into categories. The sum in quadrature of the individual uncertainties is not necessarily equal to the total
uncertainty due to correlations introduced by the fit.}
\label{tab:syst_reco}
\begin{center}
\begin{tabular}{l | c}
\toprule
& $\Delta \lca(\ttW)$ \\
\midrule
\textbf{Experimental uncertainties} & \\
Jet energy resolution &  0.013 \\
Pile-up               &  0.007 \\
$b$-tagging           &  0.005 \\
Leptons               &  0.004 \\
\MET                  &  0.004 \\
Jet energy scale      &  0.0032 \\
Luminosity            &  0.0006 \\
\midrule
\textbf{MC modelling uncertainties} & \\
\ttW modelling         &  0.013 \\
\ttZ modelling         &  0.010 \\
HF$_{e/\mu}$ modelling &  0.006 \\
\ttH modelling         &  0.005 \\
\midrule
\textbf{Other uncertainties} &  \\
$\Delta\eta^{\pm}$ CR-dependency    & 0.05 \\
\midrule
\textbf{MC statistical uncertainty} & 0.019 \\
\midrule
\textbf{Data statistical uncertainty} &  0.14 \\
\midrule
\textbf{Total uncertainty} & 0.15 \\
\bottomrule
\end{tabular}
\end{center}
\end{table}


\section{Unfolding and extraction of the charge asymmetry at particle level}
\label{sec:unfolding}

To obtain the charge asymmetry at particle level in a specific fiducial volume, an unfolding procedure is performed to correct for
detector effects, as well as for signal efficiency and acceptance effects. The procedure and the relevant definitions are explained in the following.
 
\subsection{Particle-level objects}
\label{subsec:pl-objects}
 
Particle-level objects in simulated events are defined using quasi-stable particles~(with a mean lifetime greater than \SI{30}{ps}) originating from \pp collisions.
They are selected after hadronisation but before the interaction with the various detector components or consideration of pile-up effects.
 
Particle-level electrons or muons are required to not originate from a hadron in the MC generator event record, whether directly or through a $\tau$-lepton decay. This
ensures that they originate from a \Zboson or \Wboson~boson~(where the \Wboson~boson can come either from prompt \Wboson production or a top-quark decay), without
requiring a direct match with the parent particle. The four-momenta of the bare leptons are modified~(\enquote{dressed}) by adding the four-momenta of all radiated photons
within a cone of size $\DeltaR = 0.1$, excluding photons from hadron decays, to take into account FSR photons.
 
Particle-level jets are reconstructed with the \antikt algorithm with a radius parameter of $R = 0.4$ applied to all stable particles, but excluding the neutrinos
originating from \Wboson or \Zboson~bosons and the selected electrons, muons and photons used in the definition of the charged leptons. If $b$-hadrons
with $\pt > \SI{5}{\GeV}$ are found in the MC event record, they are clustered into stable-particle jets with their energies set to negligible positive values~(referred to
as \enquote{ghost-matching})~\cite{Cacciari:2008gn}. Particle-level jets containing at least one of these $b$-hadrons are considered as $b$-jets. The particle-level missing
transverse momentum is defined as the vectorial sum of the transverse momenta of all neutrinos found in the MC simulation history of the event, excluding those originating
from hadron decays.
 
\subsection{Particle-level fiducial volume}
\label{subsec:pl-volume}
 
The particle-level fiducial volume is defined by the following requirements on the particle-level objects, as defined in Section~\ref{subsec:pl-objects}:
\begin{itemize}
\item Three electrons or muons with $\pt > \SI{15}{\GeV}$ and $\abseta < 2.5$.
\item The invariant mass of all OSSF lepton pairs has to be larger than \SI{25}{\GeV}.
\item No \Zboson-candidate~(as defined in Section~\ref{sec:regions}) among the leptons.
\item At least two jets with $\pt > \SI{20}{\GeV}$, $\abseta < 2.5$ and least one of them identified as a $b$-jet.
\end{itemize}
 
\subsection{Unfolding procedure and charge-asymmetry extraction}
 
The unfolding procedure is applied to the observed number of data events in the SRs. Analogously to the method used at detector level, described in Section~\ref{sec:leptontop},
a method of matching leptons to top quarks~(antiquarks) is required to obtain the response matrix, essential to the unfolding procedure. To reproduce the particle-level fiducial
volume, a simpler scheme is adopted that is independent of the generator-specific MC event record and any multivariate algorithm. Each lepton is combined with the closest $b$-jet
in the \DeltaR space. The $\ell$--$b$ system that yields a mass closest to the most probable mass for a $\ell$--$b$ system originating from a top-quark decay~(according to the
nominal \ttW simulation) is used to select the even and odd leptons. This procedure has an efficiency of approximately 65\% to identify the correct leptons.
 
The following formula is used for the unfolding:
\begin{equation}
\small
N_{i}^{\text{folded}} = \underbrace{ \frac{1}{\alpha_{i}} \, \sum_{j} \varepsilon_{j} M_{ij} }_{R_{ij}} N_{j}^{\text{fid}} \hspace{0.4cm} \text{with} \hspace{0.4cm} M_{ij} = \frac{ N_{ij}^{(\text{reco}\ \cap\ \text{fid})}}{N_{j}^{(\text{reco}\ \cap\ \text{fid})}}, \quad \alpha_{i} = \frac{ N_{i}^{(\text{reco}\ \cap\ \text{fid})}}{N_{i}^{\text{reco}}}, \quad \varepsilon_{j} = \frac{ N_{j}^{(\text{reco}\ \cap\ \text{fid})}}{N_{j}^{\text{fid}}},
\label{eq:unfolding_equation}
\end{equation}
with the number $N_{j}^{\text{fid}}$ representing the content of bin $j$ in the fiducial volume, $N_{i}^{\text{reco}}$ being the events in bin $i$ that satisfy the detector level selection and the symbol $\cap$ represents the logical intersection of the two regions. The $N_{i}^{\text{folded}}$ represents the bin content ($i$) after the folding of the particle level bins ($j$) has been performed through the response matrix~($R_{ij}$). This matrix is constructed
from the migration matrix~($M_{ij}$) and the acceptance and efficiency correction terms~($\alpha_{i}$ and $\varepsilon_{j}$) for each bin. The entries
in the migration matrix represent the fractions of events at particle level in a $y$-axis bin that are reconstructed at detector level in an $x$-axis bin.
They are normalised such that the sum of entries in each row is equal to one. The acceptance corrections $\alpha_{i}$ account for events that are generated
outside the fiducial volume but satisfy the selection at detector level, as described in Section~\ref{sec:regions}. The efficiency
corrections $\varepsilon_{j}$ account for events that are in the fiducial volume but fail to satisfy the detector-level selection.
 
\begin{figure}[!htpb]
\centering
\subfloat[]{\includegraphics[width=0.50\textwidth]{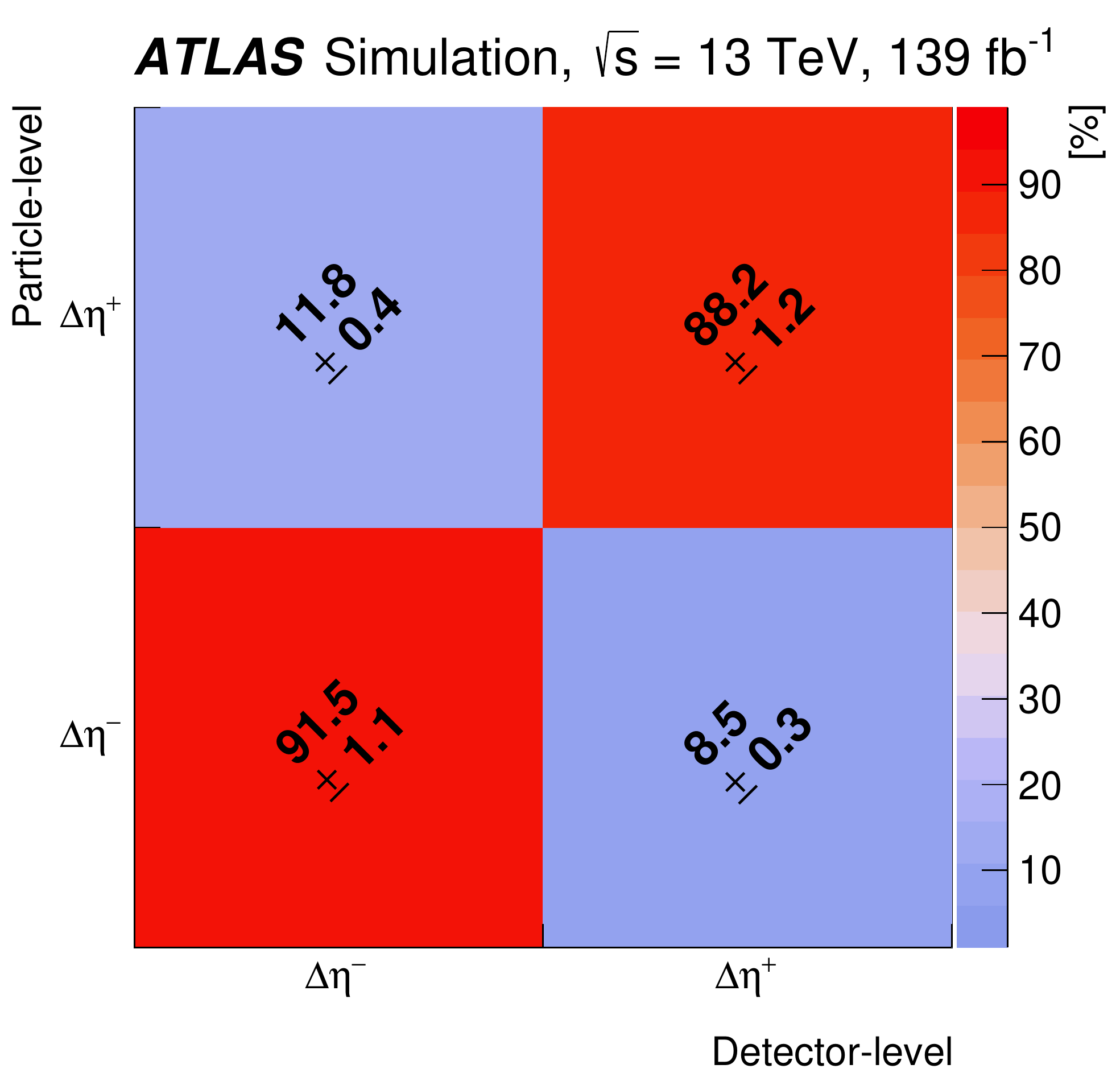}}
 
\subfloat[]{\includegraphics[width=0.49\textwidth]{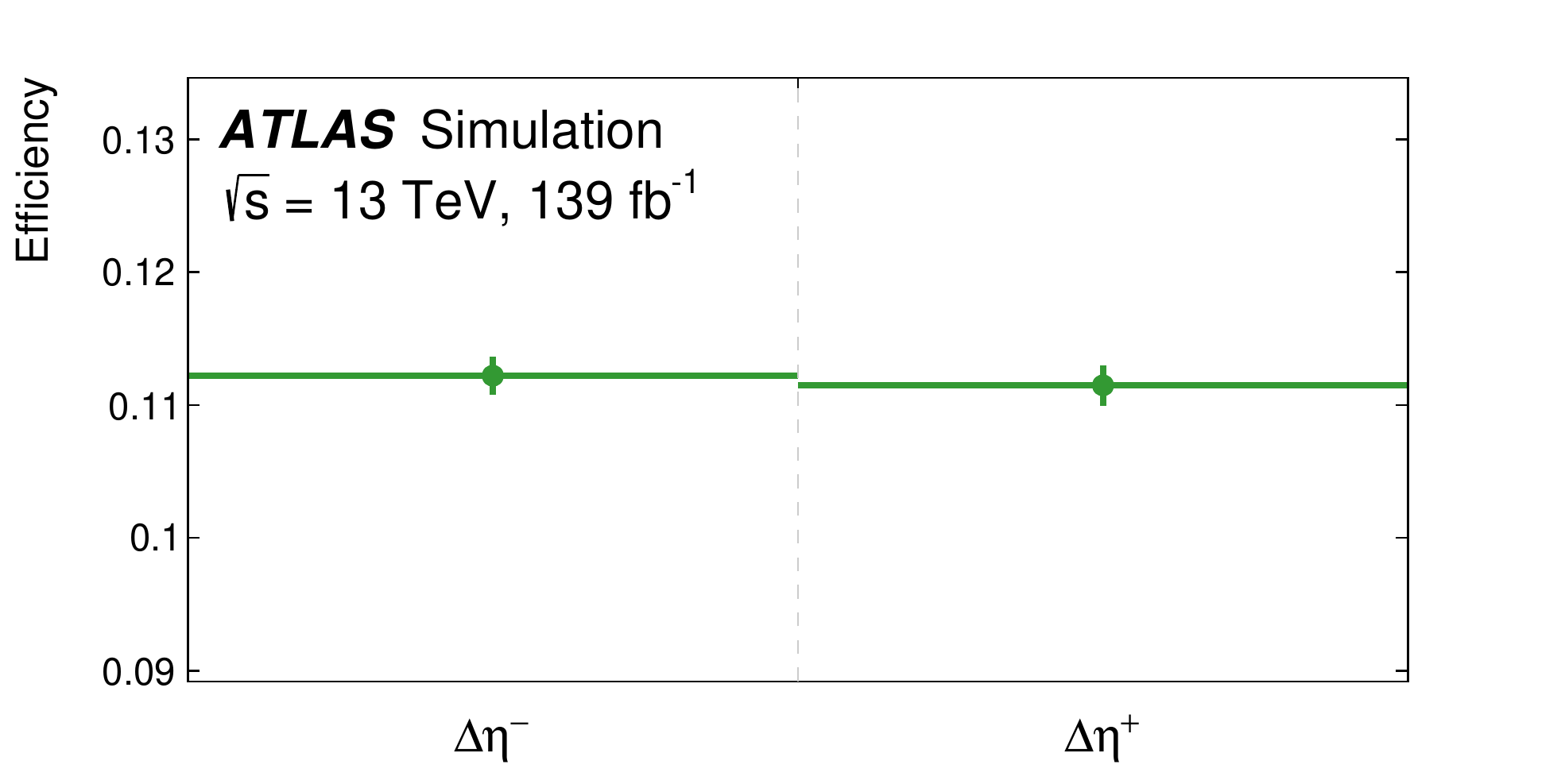}}
\subfloat[]{\includegraphics[width=0.49\textwidth]{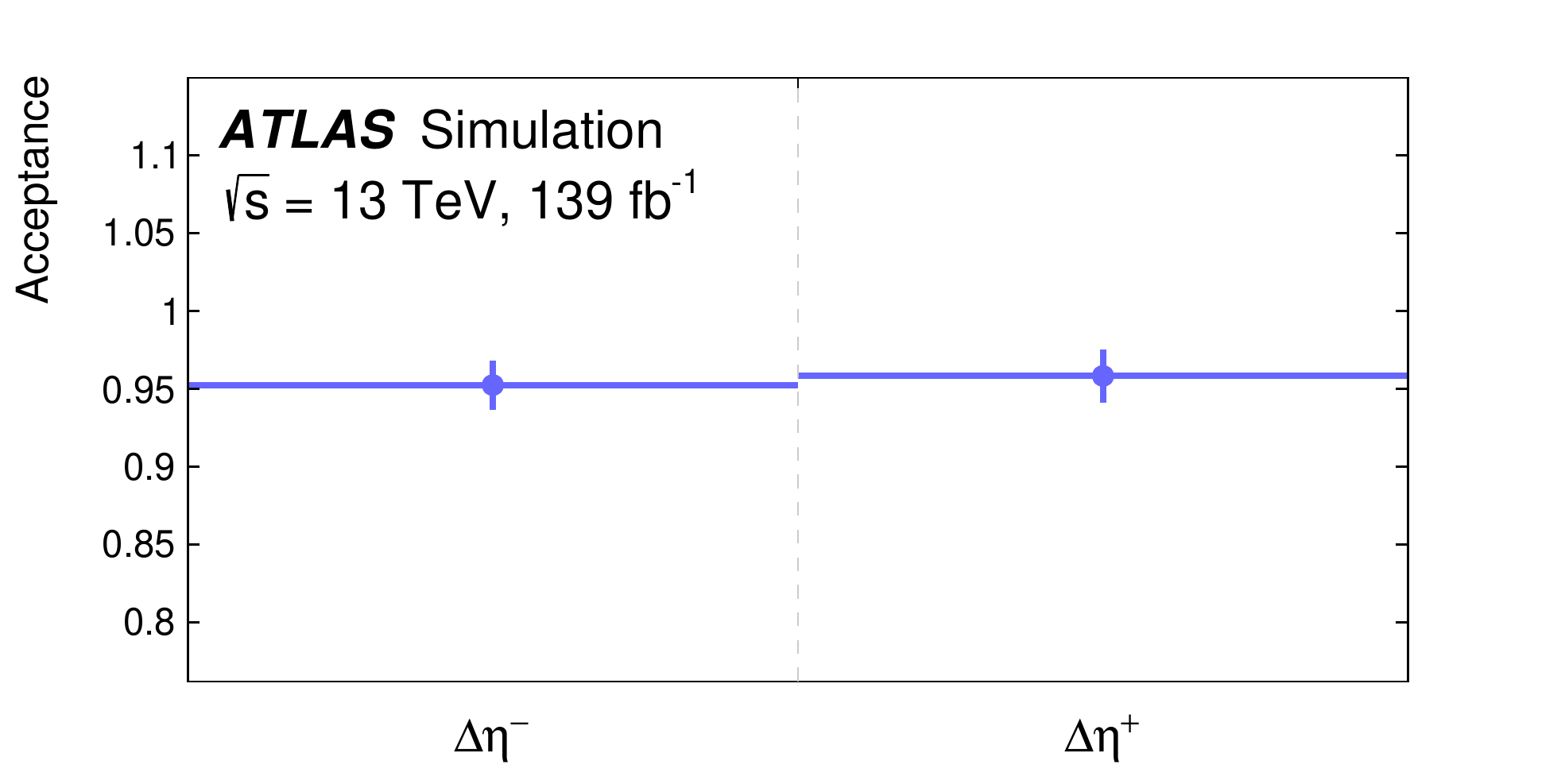}}
\caption{(a)~The migration matrix and (b,c)~the efficiency/acceptance corrections that are used as input for the unfolding of \SRIIbL. The matrices are normalised
such that the sum of any given row is 100\%, although small differences may be present due to rounding. The error bars of the efficiency/acceptance correction terms represent
the MC statistical uncertainties per bin based on the nominal \ttW \SHERPA sample.}
\label{fig:unfolding-input-SRIIbL}
\end{figure}

The migration matrices, as well as the acceptance and efficiency correction terms, are built separately for each of the SRs defined in Table~\ref{tab:RegionsCuts}.
As an example, Figure~\ref{fig:unfolding-input-SRIIbL} shows (a)~the migration matrix, as well as (b)~the efficiency and (c)~the acceptance correction factors that are used
for \SRIIbL, which is the region with the highest \ttW purity. The fraction of events in the diagonal elements of the migration matrix shows the quality of the
resolution for \dEta, which is around 90\%. The efficiency corrections are at a level of 11\%--12\% and the acceptance corrections are around 95\%. None show any notable
dependence on \dEta.
 
The unfolding procedure is the same as in Ref.~\cite{ATLAS:2022wec} and based on a profile-likelihood approach~(\enquote{profile-likelihood unfolding}). With this approach,
the unfolding problem is transformed into a standard problem of fitting normalisations of distributions. Each bin in the particle-level distribution is \enquote{folded} through the
response matrix via Eq.~(\ref{eq:unfolding_equation}), resulting in the same numbers of bins at detector level. The particle-level bins are treated as separate subsamples that
are multiplied by their respective entries in the response matrix and freely floating parameters are assigned to each of these subsamples at detector level. Analogously to the fit
described in Section~\ref{sec:results}, the freely floating parameters are assigned to the major backgrounds in the SRs: \NormttZ, \NormCO, \NormHFe and \NormHFmu. These normalisations
and the analysis regions are split into \dEtaP\ and \dEtaM, in the same way as the detector-level results. Thus, the detector-level distributions are scaled by some factors, determined
by fitting the data, and these factors are then used to scale the corresponding particle-level bins, which gives the desired unfolded result. The charge asymmetry is defined as the
parameter of interest and is related to the normalisation factors in the unfolded bins. For the CRs, no response matrices are built. However, as the signal contamination in these
regions is very small compared with the total event yields, an approximation is made whereby the signal is treated as an additional background. An exception is \CRConv where, due to
the high signal contamination, response matrices are also built. No regularisation is applied in the unfolding.
 
The systematic uncertainties in the signal and background processes considered for the unfolded results are the same as for the results at detector level~(described in
Section~\ref{sec:systematics}). Systematic uncertainties in the background processes are propagated to the unfolded distributions by varying the detector-level distributions
within their uncertainties and repeating the unfolding procedure. The modelling uncertainties of the \ttW signal are propagated through the unfolding procedure, using variations
of the response matrices.
 
An injection test is performed to verify that non-SM \lca values can be recovered in the unfolding procedure. This is done by injecting the non-SM \lca values into the particle-level
predictions, which are propagated to detector level and treated as pseudo-data in the fit. The unfolding procedure is then performed on this pseudo-data for several positive
and negative deviations from the SM \lca to compute the relation between the injected and extracted \lca values and to estimate the bias that the fit procedure introduces. After the
fit to real data, the observed \lca is substituted into the relation to extract the bias. The bias estimated from this procedure is found to be \num{0.004}. Although this value is
well covered by the systematic uncertainties, it is added as an extra uncertainty in the unfolded \lca value to account for this effect.
 
The charge asymmetry value, unfolded to particle level~(PL) in the fiducial volume defined in Section~\ref{subsec:pl-volume}, is found to be
\begin{equation*}
\lca(\ttW)^{\text{PL}} = -0.11 \pm 0.17 \,\text{(stat.)} \pm 0.05 \,\text{(syst.)},
\end{equation*}
with a SM expectation calculated using the nominal \ttW \SHERPA simulation of
\begin{equation*}
\lca(\ttW)_{\text{SM}}^{\text{PL}} = -0.063 \, ^{+0.007}_{-0.004} \,\text{(scale)} \pm 0.004 \,\text{(MC stat.)}.
\end{equation*}
The nominal values for the background normalisations are the same as reported in Section~\ref{sec:results}. The contributions from the most relevant uncertainties in the charge
asymmetry at particle level are given in Table~\ref{tab:syst_PL}. The sources of uncertainty are similar to the ones reported in Table~\ref{tab:syst_reco}, with the $\Delta\eta^{\pm}$
CR-dependency, the modelling of the \ttW and \ttZ MC processes and the statistical uncertainty being the dominant ones. The statistical uncertainty is slightly increased relative to the
detector-level result due to the unfolding procedure.
 
In principle, the \ttW \lca has sensitivity to certain linear combinations of SMEFT coefficients that parameterise the effect of operators involving four quarks. However, the constraints on the coefficients obtainable with this result are not competitive with those from \ttbar production cross--sections in the boosted regime and asymmetries~\cite{TOPQ-2019-23,TOPQ-2018-11,TOPQ-2019-28,TOPQ-2020-06}. Moreover, they are not expected to be competitive even with the large datasets of the High-Luminosity LHC.

\begin{table}[!htb]
\scriptsize
\def\arraystretch{1.2}
\caption{List of the most relevant systematic and statistical uncertainties in the leptonic charge asymmetry at particle level $\lca(\ttW)^{\text{PL}}$. For this table,
the uncertainties are symmetrised and grouped into categories. The sum in quadrature of the individual uncertainties is not necessarily equal to the total uncertainty due
to correlations introduced by the fit.}
\label{tab:syst_PL}
\begin{center}
\begin{tabular}{l | c}
\toprule
& $\Delta \lca(\ttW)^{\text{PL}}$ \\
\midrule
\textbf{Experimental uncertainties} & \\
Leptons               &  0.014 \\
Jet energy resolution &  0.011 \\
Pile-up               &  0.008 \\
Jet energy scale      &  0.004 \\
\MET                  &  0.0015 \\
Luminosity            &  0.0011 \\
Jet vertex tagger     &  0.00027 \\
\midrule
\textbf{MC modelling uncertainties} & \\
\ttW modelling         & 0.022 \\
\ttZ modelling         & 0.017 \\
HF$_{e/\mu}$ modelling & 0.015 \\
Others modelling       & 0.015 \\
\VVjets modelling      & 0.014 \\
\ttH modelling         & 0.006 \\
\midrule
\textbf{Other uncertainties} &  \\
Unfolding bias                   & 0.004 \\
$\Delta\eta^{\pm}$ CR-dependency & 0.04 \\
\midrule
\textbf{MC statistical uncertainty} &  0.027 \\
\midrule
\textbf{Response matrix}            &  0.009 \\
\midrule
\textbf{Data statistical uncertainty} & 0.17   \\
\midrule
\textbf{Total uncertainty} & 0.18 \\
\bottomrule
\end{tabular}
\end{center}
\end{table}


\section{Conclusions}
\label{sec:conclusions}

This paper presents a search for the leptonic charge asymmetry in \ttW production using \pp collision data at $\rts = \SI{13}{\TeV}$ with the full \RunTwo data sample collected
with the ATLAS detector at the LHC, corresponding to an integrated luminosity of \SI{\lumi}{\ifb}. The leptonic charge asymmetry is defined as the pseudorapidity difference between
the two reconstructed charged leptons associated with top quarks~(or top antiquarks). The search is performed in \lll final states using reconstructed light leptons~(electrons or muons),
together with jets and $b$-jets. To correctly match the leptons to either top quarks or top antiquarks, a technique based on a BDT is used.
 
The charge asymmetry at reconstruction level is obtained by performing a simultaneous profile-likelihood fit to data in different signal and control regions optimised for either
the \ttW process or the major SM background processes~(\ttZ, non-prompt leptons from HF decays or electrons from \gam-conversions). The charge asymmetry is extracted together
with the normalisations for these background processes and is found to be
\begin{equation*}
\lca(\ttW) = -0.12 \pm 0.14 \,\text{(stat.)} \pm 0.05 \,\text{(syst.)}.
\end{equation*}
This is consistent with the SM expectation of
\begin{equation*}
\lca(\ttW)_{\text{SM}} = -0.084 \, ^{+0.005}_{-0.003} \,\text{(scale)} \pm 0.006 \,\text{(MC stat.)},
\end{equation*}
calculated using the nominal \ttW \SHERPA simulation. An unfolding procedure is used to obtain the charge asymmetry at particle level in a specific fiducial volume in the \lll channel.
The unfolding is based on a profile-likelihood approach, where the unfolding is performed together with fitting normalisations of the major background processes, using the same procedure
used to derive the charge asymmetry at reconstruction level. The charge asymmetry at particle level yields
\begin{equation*}
\lca(\ttW)^{\text{PL}} = -0.11 \pm 0.17 \,\text{(stat.)} \pm 0.05 \,\text{(syst.)},
\end{equation*}
with a SM expectation calculated using the nominal \ttW \SHERPA simulation of
\begin{equation*}
\lca(\ttW)_{\text{SM}}^{\text{PL}} = -0.063 \, ^{+0.007}_{-0.004} \,\text{(scale)} \pm 0.004 \,\text{(MC stat.)}.
\end{equation*}
The most relevant systematic uncertainties affecting this search are the $\Delta\eta^{\pm}$ CR-dependency of the fit, as well as the modelling uncertainties of
the \ttW and \ttZ MC processes in the \lll channel. However, both the reconstruction- and particle-level results are severely limited by the statistical uncertainties of the data.


\clearpage
 
\section*{Acknowledgements}


We thank CERN for the very successful operation of the LHC, as well as the
support staff from our institutions without whom ATLAS could not be
operated efficiently.
 
We acknowledge the support of
ANPCyT, Argentina;
YerPhI, Armenia;
ARC, Australia;
BMWFW and FWF, Austria;
ANAS, Azerbaijan;
SSTC, Belarus;
CNPq and FAPESP, Brazil;
NSERC, NRC and CFI, Canada;
CERN;
ANID, Chile;
CAS, MOST and NSFC, China;
Minciencias, Colombia;
MEYS CR, Czech Republic;
DNRF and DNSRC, Denmark;
IN2P3-CNRS and CEA-DRF/IRFU, France;
SRNSFG, Georgia;
BMBF, HGF and MPG, Germany;
GSRI, Greece;
RGC and Hong Kong SAR, China;
ISF and Benoziyo Center, Israel;
INFN, Italy;
MEXT and JSPS, Japan;
CNRST, Morocco;
NWO, Netherlands;
RCN, Norway;
MEiN, Poland;
FCT, Portugal;
MNE/IFA, Romania;
JINR;
MES of Russia and NRC KI, Russian Federation;
MESTD, Serbia;
MSSR, Slovakia;
ARRS and MIZ\v{S}, Slovenia;
DSI/NRF, South Africa;
MICINN, Spain;
SRC and Wallenberg Foundation, Sweden;
SERI, SNSF and Cantons of Bern and Geneva, Switzerland;
MOST, Taiwan;
TENMAK, T\"urkiye;
STFC, United Kingdom;
DOE and NSF, United States of America.
In addition, individual groups and members have received support from
BCKDF, CANARIE, Compute Canada and CRC, Canada;
COST, ERC, ERDF, Horizon 2020 and Marie Sk{\l}odowska-Curie Actions, European Union;
Investissements d'Avenir Labex, Investissements d'Avenir Idex and ANR, France;
DFG and AvH Foundation, Germany;
Herakleitos, Thales and Aristeia programmes co-financed by EU-ESF and the Greek NSRF, Greece;
BSF-NSF and GIF, Israel;
Norwegian Financial Mechanism 2014-2021, Norway;
NCN and NAWA, Poland;
La Caixa Banking Foundation, CERCA Programme Generalitat de Catalunya and PROMETEO and GenT Programmes Generalitat Valenciana, Spain;
G\"{o}ran Gustafssons Stiftelse, Sweden;
The Royal Society and Leverhulme Trust, United Kingdom.
 
The crucial computing support from all WLCG partners is acknowledged gratefully, in particular from CERN, the ATLAS Tier-1 facilities at TRIUMF (Canada), NDGF (Denmark, Norway, Sweden), CC-IN2P3 (France), KIT/GridKA (Germany), INFN-CNAF (Italy), NL-T1 (Netherlands), PIC (Spain), ASGC (Taiwan), RAL (UK) and BNL (USA), the Tier-2 facilities worldwide and large non-WLCG resource providers. Major contributors of computing resources are listed in Ref.~\cite{ATL-SOFT-PUB-2021-003}.


\clearpage

\printbibliography

@article{Avoni:2018,
  author  = "G. Avoni and others",
  title   = "{The new LUCID-2 detector for luminosity measurement and monitoring in ATLAS}",
  journal = "JINST",
  volume  = "13",
  number  = "07",
  pages   = "P07017",
  doi     = "10.1088/1748-0221/13/07/P07017",
  year    = "2018"
}

@article{Alwall:2011uj-c,
      author         = "Alwall, Johan and Herquet, Michel and Maltoni, Fabio and Mattelaer, Olivier and Stelzer, Tim",
      title          = "{MadGraph 5 : going beyond}",
      journal        = "JHEP",
      volume         = "06",
      pages          = "128",
      doi            = "10.1007/JHEP06(2011)128",
      year           = "2011",
      eprint         = "1106.0522",
      archivePrefix  = "arXiv",
      primaryClass   = "hep-ph"
}

@article{Broggio:2019ewu,
      author         = "Broggio, Alessandro and Ferroglia, Andrea and Frederix, Rikkert and Pagani, Davide and Pecjak, Benjamin D. and Tsinikos, Ioannis",
      title          = "{Top-quark pair hadroproduction in association with a heavy boson at NLO+NNLL including EW corrections}",
      journal        = "JHEP",
      volume         = "08",
      year           = "2019",
      pages          = "039",
      doi            = "10.1007/JHEP08(2019)039",
      reportNumber   = "TUM-HEP-1208/19, LU-TP 19-30, IPPP/19/57",
      eprint         = "1907.04343",
      archivePrefix  = "arXiv",
      primaryClass   = "hep-ph",
}

@article{CDF:2011xdt,
    author        = "{CDF Collaboration}",
    title         = "{Evidence for a mass dependent forward-backward asymmetry in top quark pair production}",
    eprint        = "1101.0034",
    archivePrefix = "arXiv",
    primaryClass  = "hep-ex",
    reportNumber  = "FERMILAB-PUB-10-525-E",
    doi           = "10.1103/PhysRevD.83.112003",
    journal       = "Phys. Rev. D",
    volume        = "83",
    pages         = "112003",
    year          = "2011"
}

@article{CDF:2017cvy,
    author        = "{CDF and D0 Collaborations}",
    title         = "{Combined forward-backward asymmetry measurements in top-antitop quark production at the Tevatron}",
    eprint        = "1709.04894",
    archivePrefix = "arXiv",
    primaryClass = "hep-ex",
    reportNumber = "FERMILAB-PUB-17-379-E",
    doi          = "10.1103/PhysRevLett.120.042001",
    journal      = "Phys. Rev. Lett.",
    volume       = "120",
    number       = "4",
    pages        = "042001",
    year         = "2018"
}

@article{Maltoni:2014zpa,
    author        =  "Maltoni, F. and Mangano, M. L. and Tsinikos, I. and Zaro, M.",
    title         = "{Top-quark charge asymmetry and polarization in $t\overline{t}W^{\pm}$ production at the LHC}",
    eprint        = "1406.3262",
    archivePrefix = "arXiv",
    primaryClass  = "hep-ph",
    reportNumber  = "CP3-14-23, CERN-PH-TH-2014-102, LPN14-075",
    doi           = "10.1016/j.physletb.2014.07.033",
    journal       = "Phys. Lett. B",
    volume        = "736",
    pages         = "252--260",
    year          = "2014"
}

@article{Bevilacqua:2020srb,
    author        = "Bevilacqua, Giuseppe and Bi, Huan-Yu and Hartanto, Heribertus Bayu and Kraus, Manfred and Nasufi, Jasmina and Worek, Malgorzata",
    title         = "{NLO QCD corrections to off-shell ${t\bar{t}W^\pm}$ production at the LHC: correlations and asymmetries}",
    eprint        = "2012.01363",
    archivePrefix = "arXiv",
    primaryClass  = "hep-ph",
    reportNumber  = "TTK-20-43, P3H-20-078, CAVENDISH-HEP-20/14",
    doi           = "10.1140/epjc/s10052-021-09478-x",
    journal       = "Eur. Phys. J. C",
    volume        = "81",
    pages         = "675",
    year          = "2021"
}

@article{Cacciari:2008gn,
      author         = "Cacciari, Matteo and Salam, Gavin P. and Soyez, Gregory",
      title          = "{The catchment area of jets}",
      journal        = "JHEP",
      volume         = "04",
      year           = "2008",
      pages          = "005",
      doi            = "10.1088/1126-6708/2008/04/005",
      eprint         = "0802.1188",
      archivePrefix  = "arXiv",
      primaryClass   = "hep-ph",
      reportNumber   = "LPTHE-07-02",
      SLACcitation   = "%%CITATION = ARXIV:0802.1188;%%"
}

@article{Bellm:2019zci,
    author        = "Bellm, Johannes and others",
    title         = "{Herwig 7.2 release note}",
    eprint        = "1912.06509",
    archivePrefix = "arXiv",
    primaryClass  = "hep-ph",
    reportNumber  = "MAN/HEP/2019/011, CERN-TH-2019-213, IFJPAN-IV-2019-18, HERWIG-2019-02, UWTHPH-19-36, KA-TP-24-2019, LU-TP 19-57, MCnet-19-28, IPPP/19/91",
    doi           = "10.1140/epjc/s10052-020-8011-x",
    journal       = "Eur. Phys. J. C",
    volume        = "80",
    number        = "5",
    pages         = "452",
    year          = "2020"
}

@article{scikit-learn,
    title   = "{Scikit-learn: machine learning in python}",
    author  = "Pedregosa, F. and Varoquaux, G. and Gramfort, A. and Michel, V. and Thirion, B. and Grisel, O. and Blondel, M. and Prettenhofer, P. and Weiss, R. and Dubourg, V. and Vanderplas, J. and Passos, A. and  Cournapeau, D. and Brucher, M. and Perrot, M. and Duchesnay, E.",
    url     = "https://dl.acm.org/doi/10.5555/1953048.2078195",
    journal = "Journal of Machine Learning Research",
    volume  =  "12",
    pages   = "2825",
    year    = "2011"
}

@article{Workman:2022ynf,
    author        = "{Particle Data Group}",
    title         = "{Review of Particle Physics}",
    doi           = "10.1093/ptep/ptac097",
    journal       = "PTEP",
    volume        = "2022",
    pages         = "083C01",
    year          = "2022"
}

@article{Grzadkowski:2010es,
    author         = "Grzadkowski, B. and Iskrzynski, M. and Misiak, M. and Rosiek, J.",
    title          =  "{Dimension-six terms in the Standard Model Lagrangian}",
    eprint         = "1008.4884",
    archivePrefix  = "arXiv",
    primaryClass   = "hep-ph",
    reportNumber   = "IFT-9-2010, TTP10-35",
    doi            = "10.1007/JHEP10(2010)085",
    journal        = "JHEP",
    volume         = "10",
    pages          = "085",
    year           = "2010"
}

@article{CMS:2022tkv,
    author        = "{CMS Collaboration}",
    title         = "{Measurement of the cross section of top quark-antiquark pair production in association with a W boson in proton-proton collisions at $\sqrt{s} = 13\,\text{TeV}$}",
    eprint        = "2208.06485",
    archivePrefix = "arXiv",
    primaryClass  = "hep-ex",
    reportNumber  = "CMS-TOP-21-011, CERN-EP-2022-168",
    month         = "8",
    year          = "2022"
}

@article{CMS:2022ged,
    author        = "{CMS Collaboration}",
    title         = "{Measurement of the $t\bar{t}$ charge asymmetry in events with highly Lorentz-boosted top quarks in $pp$ collisions at $\sqrt{s} = 13\,\text{TeV}$}",
    eprint        = "2208.02751",
    archivePrefix = "arXiv",
    primaryClass  = "hep-ex",
    reportNumber  = "CMS-TOP-21-014, CERN-EP-2022-155",
    month         = "8",
    year          = "2022"
}

@article{Brivio:2017vri,
    author        = "Brivio, Ilaria and Trott, Michael",
    title         = "{The Standard Model as an Effective Field Theory}",
    eprint        = "1706.08945",
    archivePrefix = "arXiv",
    primaryClass  = "hep-ph",
    doi           = "10.1016/j.physrep.2018.11.002",
    journal       = "Phys. Rept.",
    volume        = "793",
    pages         = "1--98",
    year          = "2019"
}

@article{ATLAS:2022wec,
    author        = "{ATLAS Collaboration}",
    title         = "{Measurement of the charge asymmetry in top-quark pair production in association with a photon with the ATLAS experiment}",
    eprint        = "2212.10552",
    archivePrefix = "arXiv",
    primaryClass  = "hep-ex",
    reportNumber  = "CERN-EP-2022-260",
    month         = "12",
    year          = "2022"
}

@article{FTAG-2019-07,
    author         = "{ATLAS Collaboration}",
    title          = "{ATLAS flavour-tagging algorithms for the LHC Run 2 $pp$ collision dataset}",
    eprint         = "2211.16345",
    archivePrefix  = "arXiv",
    primaryClass   = "physics.data-an",
    reportNumber   = "CERN-EP-2022-226",
    month          = "11",
    year           = "2022"
}

@article{ATLAS:2019bsc,
    author         = "{ATLAS Collaboration}",
    title = "{Measurement of $W^{\pm}Z$ production cross sections and gauge boson polarisation in $pp$ collisions at $\sqrt{s} = 13$ TeV with the ATLAS detector}",
    eprint = "1902.05759",
    archivePrefix = "arXiv",
    primaryClass = "hep-ex",
    reportNumber = "CERN-EP-2018-327",
    doi = "10.1140/epjc/s10052-019-7027-6",
    journal = "Eur. Phys. J. C",
    volume = "79",
    number = "6",
    pages = "535",
    year = "2019"
}

@article{ATLAS:2023ajo,
    author = "{ATLAS Collaboration}",
    title = "{Observation of four-top-quark production in the multilepton final state with the ATLAS detector}",
    eprint = "2303.15061",
    archivePrefix = "arXiv",
    primaryClass = "hep-ex",
    reportNumber = "CERN-EP-2023-055",
    month = "3",
    year = "2023"
}

@Article{Cacciari:2008gp,
     author    = "Cacciari, Matteo and Salam, Gavin P. and Soyez, Gregory",
     title     = "{The anti-\(k_{t}\) jet clustering algorithm}",
     journal   = "JHEP",
     volume    = "04",
     year      = "2008",
     pages     = "063",
     eprint    = "0802.1189",
     archivePrefix = "arXiv",
     primaryClass  =  "hep-ph",
     doi       = "10.1088/1126-6708/2008/04/063",
     SLACcitation  = "%%CITATION = 0802.1189;%%"
}

@Article{Fastjet,
      author         = "Cacciari, Matteo and Salam, Gavin P. and Soyez, Gregory",
      title          = "{FastJet user manual}",
      journal        = "Eur. Phys. J. C",
      volume         = "72",
      year           = "2012",
      pages          = "1896",
      doi            = "10.1140/epjc/s10052-012-1896-2",
      eprint         = "1111.6097",
      archivePrefix  = "arXiv",
      primaryClass   = "hep-ph",
      reportNumber   = "CERN-PH-TH-2011-297",
      SLACcitation   = "%%CITATION = ARXIV:1111.6097;%%"
}

@Article{Butterworth:2015oua,
      author         = "Butterworth, Jon and others",
      title          = "{PDF4LHC recommendations for LHC Run II}",
      journal        = "J. Phys. G",
      volume         = "43",
      year           = "2016",
      pages          = "023001",
      doi            = "10.1088/0954-3899/43/2/023001",
      eprint         = "1510.03865",
      archivePrefix  = "arXiv",
      primaryClass   = "hep-ph",
      reportNumber   = "OUTP-15-17P, SMU-HEP-15-12, TIF-UNIMI-2015-14,
                        LCTS-2015-27, CERN-PH-TH-2015-249",
      SLACcitation   = "%%CITATION = ARXIV:1510.03865;%%"
}

@article{Ball:2012cx,
      author         = "Ball, Richard D. and others",
      title          = "{Parton distributions with LHC data}",
      journal        = "Nucl. Phys. B",
      volume         = "867",
      year           = "2013",
      pages          = "244",
      doi            = "10.1016/j.nuclphysb.2012.10.003",
      eprint         = "1207.1303",
      archivePrefix  = "arXiv",
      primaryClass   = "hep-ph",
      reportNumber   = "EDINBURGH-2012-08, IFUM-FT-997, FR-PHENO-2012-014,
                        RWTH-TTK-12-25, CERN-PH-TH-2012-037, SFB-CPP-12-47\,
                        --CERN-PH-TH-2012-037",
      SLACcitation   = "%%CITATION = ARXIV:1207.1303;%%"
}

@article{Ball:2014uwa,
      author         = "Ball, Richard D. and others",
      title          = "{Parton distributions for the LHC run II}",
      collaboration  = "NNPDF",
      journal        = "JHEP",
      volume         = "04",
      year           = "2015",
      pages          = "040",
      doi            = "10.1007/JHEP04(2015)040",
      eprint         = "1410.8849",
      archivePrefix  = "arXiv",
      primaryClass   = "hep-ph",
      reportNumber   = "EDINBURGH-2014-15, IFUM-1034-FT, CERN-PH-TH-2013-253,
                        OUTP-14-11P, CAVENDISH-HEP-14-11",
      SLACcitation   = "%%CITATION = ARXIV:1410.8849;%%"
}

@Article{Melnikov:2006kv,
     author    = "Melnikov, Kirill and Petriello, Frank",
     title     = "{Electroweak gauge boson production at hadron colliders
                  through \(\mathcal{O}(\alpha_{s}^{2})\)}",
     journal   = "Phys. Rev. D",
     volume    = "74",
     year      = "2006",
     pages     = "114017",
     eprint    = "hep-ph/0609070",
     archivePrefix = "arXiv",
     doi       = "10.1103/PhysRevD.74.114017",
     SLACcitation  = "%%CITATION = HEP-PH/0609070;%%"
}

@article{Melnikov:2011ta,
      author         = "Melnikov, Kirill and Schulze, Markus and Scharf, Andreas",
      title          = "{QCD corrections to top quark pair production in
                        association with a photon at hadron colliders}",
      journal        = "Phys. Rev. D",
      volume         = "83",
      year           = "2011",
      pages          = "074013",
      doi            = "10.1103/PhysRevD.83.074013",
      eprint         = "1102.1967",
      archivePrefix  = "arXiv",
      primaryClass   = "hep-ph",
      SLACcitation   = "%%CITATION = ARXIV:1102.1967;%%"
}

@article{Czakon:2012pz,
      author         = "Czakon, Michal and Mitov, Alexander",
      title          = "{NNLO corrections to top pair production at hadron
                        colliders: the quark-gluon reaction}",
      journal        = "JHEP",
      volume         = "01",
      pages          = "080",
      doi            = "10.1007/JHEP01(2013)080",
      year           = "2013",
      eprint         = "1210.6832",
      archivePrefix  = "arXiv",
      primaryClass   = "hep-ph",
      SLACcitation   = "%%CITATION = ARXIV:1210.6832;%%",
}

@Article{Sjostrand:2007gs,
	author    = "Sj{\"o}strand, T. and Mrenna, S. and Skands, P.",
	title     = "{A brief introduction to PYTHIA 8.1}",
	journal   = "Comput. Phys. Commun.",
	volume    = "178",
	year      = "2008",
	pages     = "852-867",
	eprint    = "0710.3820",
	archivePrefix = "arXiv",
	primaryClass  =  "hep-ph",
	doi       = "10.1016/j.cpc.2008.01.036",
	SLACcitation  = "%%CITATION = 0710.3820;%%"
}

@article{Sjostrand:2014zea,
      author         = "Sj{\"o}strand, Torbj{\"o}rn and Ask, Stefan and Christiansen,
                        Jesper R. and Corke, Richard and Desai, Nishita and Ilten,
                        Philip and Mrenna, Stephen and Prestel, Stefan and
                        Rasmussen, Christine O. and Skands, Peter Z.",
      title          = "{An introduction to PYTHIA 8.2}",
      journal        = "Comput. Phys. Commun.",
      volume         = "191",
      year           = "2015",
      pages          = "159",
      doi            = "10.1016/j.cpc.2015.01.024",
      eprint         = "1410.3012",
      archivePrefix  = "arXiv",
      primaryClass   = "hep-ph",
      reportNumber   = "LU-TP-14-36, MCNET-14-22, CERN-PH-TH-2014-190,
                        FERMILAB-PUB-14-316-CD, DESY-14-178, SLAC-PUB-16122,
                        --FERMILAB-PUB-14-316-CD",
      SLACcitation   = "%%CITATION = ARXIV:1410.3012;%%"
}

@Article{Lange:2001uf,
      author         = "Lange, D. J.",
      title          = "{The EvtGen particle decay simulation package}",
      booktitle      = "{Proceedings, 7th International Conference on B physics
                        at hadron machines (BEAUTY 2000)}",
      journal        = "Nucl. Instrum. Meth. A",
      volume         = "462",
      year           = "2001",
      pages          = "152",
      doi            = "10.1016/S0168-9002(01)00089-4",
      SLACcitation   = "%%CITATION = NUIMA,A462,152;%%"
}

@article{Golonka:2005pn,
      author         = "Golonka, Piotr and Was, Zbigniew",
      title          = "{PHOTOS Monte Carlo: a precision tool for QED corrections
                        in \(Z\) and \(W\) decays}",
      journal        = "Eur. Phys. J. C",
      volume         = "45",
      year           = "2006",
      pages          = "97-107",
      doi            = "10.1140/epjc/s2005-02396-4",
      eprint         = "hep-ph/0506026",
      archivePrefix  = "arXiv",
      XprimaryClass   = "hep-ph",
      reportNumber   = "IFJPAN-V-05-01, CERN-PH-TH-2005-091",
      SLACcitation   = "%%CITATION = HEP-PH/0506026;%%"
}

@Article{Alwall:2014hca,
      author         = "Alwall, J. and Frederix, R. and Frixione, S. and Hirschi,
                        V. and Maltoni, F. and Mattelaer, O. and Shao, H. -S. and
                        Stelzer, T. and Torrielli, P. and Zaro, M.",
      title          = "{The automated computation of tree-level and
                        next-to-leading order differential cross sections, and
                        their matching to parton shower simulations}",
      journal        = "JHEP",
      volume         = "07",
      year           = "2014",
      pages          = "079",
      doi            = "10.1007/JHEP07(2014)079",
      eprint         = "1405.0301",
      archivePrefix  = "arXiv",
      primaryClass   = "hep-ph",
      reportNumber   = "CERN-PH-TH-2014-064, CP3-14-18, LPN14-066, MCNET-14-09,
                        ZU-TH-14-14",
      SLACcitation   = "%%CITATION = ARXIV:1405.0301;%%"
}

@article{Frederix:2012ps,
      author         = "Frederix, Rikkert and Frixione, Stefano",
      title          = "{Merging meets matching in MC@NLO}",
      journal        = "JHEP",
      volume         = "12",
      year           = "2012",
      pages          = "061",
      doi            = "10.1007/JHEP12(2012)061",
      eprint         = "1209.6215",
      archivePrefix  = "arXiv",
      primaryClass   = "hep-ph",
      reportNumber   = "CERN-PH-TH-2012-247, ZU-TH-21-12",
      SLACcitation   = "%%CITATION = ARXIV:1209.6215;%%"
}

@article{Artoisenet:2012st,
      author         = "Artoisenet, Pierre and Frederix, Rikkert and Mattelaer,
                        Olivier and Rietkerk, Robbert",
      title          = "{Automatic spin-entangled decays of heavy resonances in
                        Monte Carlo simulations}",
      journal        = "JHEP",
      volume         = "03",
      pages          = "015",
      doi            = "10.1007/JHEP03(2013)015",
      year           = "2013",
      eprint         = "1212.3460",
      archivePrefix  = "arXiv",
      primaryClass   = "hep-ph",
      reportNumber   = "NIKHEF-2012-021, CERN-PH-TH-2012-329",
      SLACcitation   = "%%CITATION = ARXIV:1212.3460;%%",
}

@article{Frixione:2007zp,
      author         = "Frixione, Stefano and Laenen, Eric and Motylinski,
                        Patrick and Webber, Bryan R.",
      title          = "{Angular correlations of lepton pairs from vector boson
                        and top quark decays in Monte Carlo simulations}",
      journal        = "JHEP",
      volume         = "04",
      year           = "2007",
      pages          = "081",
      doi            = "10.1088/1126-6708/2007/04/081",
      eprint         = "hep-ph/0702198",
      archivePrefix  = "arXiv",
      OPTprimaryClass   = "HEP-PH",
      reportNumber   = "CAVENDISH-HEP-07-01, GEF-TH-09-2007, ITP-UU-07-10,
                        NIKHEF-2007-004",
      SLACcitation   = "%%CITATION = HEP-PH/0702198;%%"
}

@Article{Corcella:2000bw,
      author         = "Corcella, G. and others",
      title          = "{HERWIG 6: an event generator for hadron emission
                        reactions with interfering gluons (including supersymmetric processes)}",
      journal        = "JHEP",
      volume         = "01",
      pages          = "010",
      doi            = "10.1088/1126-6708/2001/01/010",
      year           = "2001",
      eprint         = "hep-ph/0011363",
      archivePrefix  = "arXiv",
      reportNumber   = "CAVENDISH-HEP-99-03, CERN-TH-2000-284, RAL-TR-2000-048",
      SLACcitation   = "%%CITATION = HEP-PH/0011363;%%",
}

@Article{Bahr:2008pv,
      author         = "B{\"a}hr, M. and others",
      title          = "{Herwig++ physics and manual}",
      journal        = "Eur. Phys. J. C",
      volume         = "58",
      year           = "2008",
      pages          = "639",
      doi            = "10.1140/epjc/s10052-008-0798-9",
      eprint         = "0803.0883",
      archivePrefix  = "arXiv",
      primaryClass   = "hep-ph",
      reportNumber   = "CERN-PH-TH-2008-038, CAVENDISH-HEP-08-03, KA-TP-05-2008,
                        DCPT-08-22, IPPP-08-11, CP3-08-05",
      SLACcitation   = "%%CITATION = ARXIV:0803.0883;%%"
}

@article{Bellm:2017jjp,
      author         = "Bellm, Johannes and others",
      title          = "{Herwig 7.1 Release Note}",
      eprint         = "1705.06919",
      year           = "2017",
      archivePrefix  = "arXiv",
      primaryClass   = "hep-ph",
      reportNumber   = "CERN-PH-TH-2017-109, MAN/HEP/2017/08, UWTHPH-2017-10, 
                        IFJPAN-IV-2017-7, NIKHEF 2017-026, HERWIG-2017-02, 
                        KA-TP-19-2017, MCnet-17-08, IPPP/17/40",
      SLACcitation   = "%%CITATION = ARXIV:1705.06919;%%"
}

@Article{Frixione:2007vw,
      author         = "Frixione, Stefano and Nason, Paolo and Oleari, Carlo",
      title          = "{Matching NLO QCD computations with parton shower
                        simulations: the POWHEG method}",
      journal        = "JHEP",
      volume         = "11",
      pages          = "070",
      doi            = "10.1088/1126-6708/2007/11/070",
      year           = "2007",
      eprint         = "0709.2092",
      archivePrefix  = "arXiv",
      primaryClass   = "hep-ph",
}

@article{Hamilton:2012np,
      author         = "Hamilton, Keith and Nason, Paolo and Zanderighi, Giulia",
      title          = "{MINLO: multi-scale improved NLO}",
      journal        = "JHEP",
      volume         = "10",
      year           = "2012",
      pages          = "155",
      doi            = "10.1007/JHEP10(2012)155",
      eprint         = "1206.3572",
      archivePrefix  = "arXiv",
      primaryClass   = "hep-ph",
      reportNumber   = "CERN-PH-TH-2012-166, OUTP-12-11P, MCNET-12-07",
      SLACcitation   = "%%CITATION = ARXIV:1206.3572;%%"
}

@article{Kidonakis:2010ux,
      author         = "Kidonakis, Nikolaos",
      title          = "{Two-loop soft anomalous dimensions for single top quark
                        associated production with a \(W^{-}\) or \(H^{-}\)}",
      journal        = "Phys. Rev. D",
      volume         = "82",
      year           = "2010",
      pages          = "054018",
      doi            = "10.1103/PhysRevD.82.054018",
      eprint         = "1005.4451",
      archivePrefix  = "arXiv",
      primaryClass   = "hep-ph",
      SLACcitation   = "%%CITATION = ARXIV:1005.4451;%%"
}

@article{deFlorian:2016spz,
      author         = "de Florian, D. and others",
      title          = "{Handbook of LHC Higgs Cross Sections: 4. Deciphering the
                        Nature of the Higgs Sector}",
      collaboration  = "LHC Higgs Cross Section Working Group",
      doi            = "10.23731/CYRM-2017-002",
      year           = "2016",
      eprint         = "1610.07922",
      archivePrefix  = "arXiv",
      primaryClass   = "hep-ph",
      reportNumber   = "FERMILAB-FN-1025-T, CERN-2017-002-M",
      SLACcitation   = "%%CITATION = ARXIV:1610.07922;%%"
}

@Article{Gleisberg:2008fv,
      author         = "Gleisberg, Tanju and H{\"o}che, Stefan",
      title          = "{Comix, a new matrix element generator}",
      journal        = "JHEP",
      volume         = "12",
      pages          = "039",
      doi            = "10.1088/1126-6708/2008/12/039",
      year           = "2008",
      eprint         = "0808.3674",
      archivePrefix  = "arXiv",
      primaryClass   = "hep-ph",
}

@Article{Schumann:2007mg,
      author         = "Schumann, Steffen and Krauss, Frank",
      title          = "{A parton shower algorithm based on Catani--Seymour dipole factorisation}",
      journal        = "JHEP",
      volume         = "03",
      pages          = "038",
      doi            = "10.1088/1126-6708/2008/03/038",
      year           = "2008",
      eprint         = "0709.1027",
      archivePrefix  = "arXiv",
      primaryClass   = "hep-ph",
}

@article{Bothmann:2019yzt,
      author = "Bothmann, Enrico and others",
      title = "{Event generation with Sherpa 2.2}",
      journal = "SciPost Phys.",
      volume = "7",
      year = "2019",
      number = "3",
      pages = "034",
      doi = "10.21468/SciPostPhys.7.3.034",
      reportNumber = "FERMILAB-PUB-19-218-T, SLAC-PUB-17433, IPPP/19/42, MCNET-19-11",
      eprint = "1905.09127",
      archivePrefix = "arXiv",
      primaryClass = "hep-ph",
}

@article{Catani:2001cc,
      author         = "Catani, S. and Krauss, F. and Webber, B. R. and Kuhn, R.",
      title          = "{QCD Matrix Elements + Parton Showers}",
      journal        = "JHEP",
      volume         = "11",
      year           = "2001",
      pages          = "063",
      doi            = "10.1088/1126-6708/2001/11/063",
      eprint         = "hep-ph/0109231",
      archivePrefix  = "arXiv",
      reportNumber   = "CERN-TH-2000-367, CAVENDISH-HEP-00-03",
      SLACcitation   = "%%CITATION = HEP-PH/0109231;%%"
}

@article{Cascioli:2011va,
      author         = "Cascioli, Fabio and Maierh{\"o}fer, Philipp and Pozzorini, Stefano",
      title          = "{Scattering Amplitudes with Open Loops}",
      journal        = "Phys. Rev. Lett.",
      volume         = "108",
      year           = "2012",
      pages          = "111601",
      doi            = "10.1103/PhysRevLett.108.111601",
      eprint         = "1111.5206",
      archivePrefix  = "arXiv",
      primaryClass   = "hep-ph",
      reportNumber   = "ZU-TH-23-11, LPN11-66",
      SLACcitation   = "%%CITATION = ARXIV:1111.5206;%%"
}

@article{Denner:2016kdg,
      author         = "Denner, Ansgar and Dittmaier, Stefan and Hofer, Lars",
      title          = "{\textsc{Collier}: A fortran-based complex one-loop library in
                        extended regularizations}",
      journal        = "Comput. Phys. Commun.",
      volume         = "212",
      year           = "2017",
      pages          = "220-238",
      doi            = "10.1016/j.cpc.2016.10.013",
      eprint         = "1604.06792",
      archivePrefix  = "arXiv",
      primaryClass   = "hep-ph",
      reportNumber   = "FR-PHENO-2016-003, ICCUB-16-016",
      SLACcitation   = "%%CITATION = ARXIV:1604.06792;%%"
}

@article{Kallweit:2015dum,
    author = "Kallweit, Stefan and Lindert, Jonas M. and Maierh{\"o}fer, Philipp and Pozzorini, Stefano and Sch{\"o}nherr, Marek",
    title = "{NLO QCD+EW predictions for V + jets including off-shell vector-boson decays and multijet merging}",
    eprint = "1511.08692",
    archivePrefix = "arXiv",
    primaryClass = "hep-ph",
    reportNumber = "DCPT-15-140, FR-PHENO-2015-014, IPPP-15-70, MCNET-15-23, ZU-TH-41-15, MITP-15-108",
    doi = "10.1007/JHEP04(2016)021",
    journal = "JHEP",
    volume = "04",
    pages = "021",
    year = "2016"
}

@article{Demartin:2016axk,
    author = "Demartin, Federico and Maier, Benedikt and Maltoni, Fabio and Mawatari, Kentarou and Zaro, Marco",
    title = "{tWH associated production at the LHC}",
    eprint = "1607.05862",
    archivePrefix = "arXiv",
    primaryClass = "hep-ph",
    reportNumber = "MCNET-16-30, CP3-16-40, LPSC16158",
    doi = "10.1140/epjc/s10052-017-4601-7",
    journal = "Eur. Phys. J. C",
    volume = "77",
    number = "1",
    pages = "34",
    year = "2017"
}

@article{Agostinelli:2002hh,
      author         = "{GEANT4 Collaboration} and Agostinelli, S. and others",
      title          = "{\textsc{Geant4} -- a simulation toolkit}",
      journal        = "Nucl. Instrum. Meth. A",
      volume         = "506",
      year           = "2003",
      pages          = "250",
      doi            = "10.1016/S0168-9002(03)01368-8",
      reportNumber   = "SLAC-PUB-9350, FERMILAB-PUB-03-339",
      SLACcitation   = "%%CITATION = NUIMA,A506,250;%%"
}

@Booklet{RooFit,
      author         = "Verkerke, Wouter and Kirkby, David",
      title          = "{The RooFit toolkit for data modeling}",
      year           = "2003",
      eprint         = "physics/0306116",
      archivePrefix  = "arXiv",
      primaryClass   = "physics.data-an",
}

@Article{PERF-2007-01,
    author         = "{ATLAS Collaboration}",
    title          = "{The ATLAS Experiment at the CERN Large Hadron Collider}",
    journal        = "JINST",
    volume         = "3",
    year           = "2008",
    pages          = "S08003",
    doi            = "10.1088/1748-0221/3/08/S08003",
    primaryClass   = "hep-ex",
}

@Article{SOFT-2010-01,
    author         = "{ATLAS Collaboration}",
    title          = "{The ATLAS Simulation Infrastructure}",
    journal        = "Eur. Phys. J. C",
    volume         = "70",
    year           = "2010",
    pages          = "823",
    doi            = "10.1140/epjc/s10052-010-1429-9",
    eprint         = "1005.4568",
    archivePrefix  = "arXiv",
    primaryClass   = "physics.ins-det",
}

@Article{STDM-2012-23,
    author         = "{ATLAS Collaboration}",
    title          = "{Measurement of the \(Z/\gamma^*\) boson transverse momentum distribution in \(pp\) collisions at \(\sqrt{s} = 7\,\text{TeV}\) with the ATLAS detector}",
    journal        = "JHEP",
    volume         = "09",
    year           = "2014",
    pages          = "145",
    doi            = "10.1007/JHEP09(2014)145",
    reportNumber   = "CERN-PH-EP-2014-075",
    eprint         = "1406.3660",
    archivePrefix  = "arXiv",
    primaryClass   = "hep-ex",
}

@Article{PERF-2014-03,
    author         = "{ATLAS Collaboration}",
    title          = "{Performance of pile-up mitigation techniques for jets in \(pp\) collisions at \(\sqrt{s} = 8\,\text{TeV}\) using the ATLAS detector}",
    journal        = "Eur. Phys. J. C",
    volume         = "76",
    year           = "2016",
    pages          = "581",
    doi            = "10.1140/epjc/s10052-016-4395-z",
    reportNumber   = "CERN-PH-EP-2015-206",
    eprint         = "1510.03823",
    archivePrefix  = "arXiv",
    primaryClass   = "hep-ex",
}

@Article{PERF-2015-09,
    author         = "{ATLAS Collaboration}",
    title          = "{Jet reconstruction and performance using particle flow with the ATLAS Detector}",
    journal        = "Eur. Phys. J. C",
    volume         = "77",
    year           = "2017",
    pages          = "466",
    doi            = "10.1140/epjc/s10052-017-5031-2",
    reportNumber   = "CERN-EP-2017-024",
    eprint         = "1703.10485",
    archivePrefix  = "arXiv",
    primaryClass   = "hep-ex",
}

@Article{PERF-2015-10,
    author         = "{ATLAS Collaboration}",
    title          = "{Muon reconstruction performance of the ATLAS detector in proton--proton collision data at \(\sqrt{s} = 13\,\text{TeV}\)}",
    journal        = "Eur. Phys. J. C",
    volume         = "76",
    year           = "2016",
    pages          = "292",
    doi            = "10.1140/epjc/s10052-016-4120-y",
    reportNumber   = "CERN-EP-2016-033",
    eprint         = "1603.05598",
    archivePrefix  = "arXiv",
    primaryClass   = "hep-ex",
}

@Article{EXOT-2016-16,
    author         = "{ATLAS Collaboration}",
    title          = "{Search for new phenomena in events with same-charge leptons and \(b\)-jets in \(pp\) collisions at \(\sqrt{s} = 13\,\text{TeV}\) with the ATLAS detector}",
    journal        = "JHEP",
    volume         = "12",
    year           = "2018",
    pages          = "039",
    doi            = "10.1007/JHEP12(2018)039",
    reportNumber   = "CERN-EP-2018-171",
    eprint         = "1807.11883",
    archivePrefix  = "arXiv",
    primaryClass   = "hep-ex",
}

@Article{PERF-2016-07,
    author         = "{ATLAS Collaboration}",
    title          = "{Performance of missing transverse momentum reconstruction with the ATLAS detector using proton--proton collisions at \(\sqrt{s} = 13\,\text{TeV}\)}",
    journal        = "Eur. Phys. J. C",
    volume         = "78",
    year           = "2018",
    pages          = "903",
    doi            = "10.1140/epjc/s10052-018-6288-9",
    reportNumber   = "CERN-EP-2017-274",
    eprint         = "1802.08168",
    archivePrefix  = "arXiv",
    primaryClass   = "hep-ex",
}

@Article{TOPQ-2016-11,
    author         = "{ATLAS Collaboration}",
    title          = "{Measurement of the \(t\bar{t}Z\) and \(t\bar{t}W\) cross sections in proton--proton collisions at \(\sqrt{s} = 13\,\text{TeV}\) with the ATLAS detector}",
    journal        = "Phys. Rev. D",
    volume         = "99",
    year           = "2019",
    pages          = "072009",
    doi            = "10.1103/PhysRevD.99.072009",
    reportNumber   = "CERN-EP-2018-331",
    eprint         = "1901.03584",
    archivePrefix  = "arXiv",
    primaryClass   = "hep-ex",
}

@Article{TOPQ-2016-16,
    author         = "{ATLAS and CMS Collaborations}",
    title          = "{Combination of inclusive and differential \(t\bar{t}\) charge asymmetry measurements using ATLAS and CMS data at \(\sqrt{s} = 7\) and \(8\,\text{TeV}\)}",
    journal        = "JHEP",
    volume         = "04",
    year           = "2018",
    pages          = "033",
    doi            = "10.1007/JHEP04(2018)033",
    eprint         = "1709.05327",
    archivePrefix  = "arXiv",
    primaryClass   = "hep-ex",
}

@Article{TRIG-2016-01,
    author         = "{ATLAS Collaboration}",
    title          = "{Performance of the ATLAS trigger system in 2015}",
    journal        = "Eur. Phys. J. C",
    volume         = "77",
    year           = "2017",
    pages          = "317",
    doi            = "10.1140/epjc/s10052-017-4852-3",
    reportNumber   = "CERN-EP-2016-241",
    eprint         = "1611.09661",
    archivePrefix  = "arXiv",
    primaryClass   = "hep-ex",
}

@Article{HIGG-2017-02,
    author         = "{ATLAS Collaboration}",
    title          = "{Evidence for the associated production of the Higgs boson and a top quark pair with the ATLAS detector}",
    journal        = "Phys. Rev. D",
    volume         = "97",
    year           = "2018",
    pages          = "072003",
    doi            = "10.1103/PhysRevD.97.072003",
    reportNumber   = "CERN-EP-2017-281",
    eprint         = "1712.08891",
    archivePrefix  = "arXiv",
    primaryClass   = "hep-ex",
}

@Article{PERF-2017-01,
    author         = "{ATLAS Collaboration}",
    title          = "{Electron reconstruction and identification in the ATLAS experiment using the 2015 and 2016 LHC proton--proton collision data at \(\sqrt{s} = 13\,\text{TeV}\)}",
    journal        = "Eur. Phys. J. C",
    volume         = "79",
    year           = "2019",
    pages          = "639",
    doi            = "10.1140/epjc/s10052-019-7140-6",
    reportNumber   = "CERN-EP-2018-273",
    eprint         = "1902.04655",
    archivePrefix  = "arXiv",
    primaryClass   = "hep-ex",
}

@Article{EGAM-2018-01,
    author         = "{ATLAS Collaboration}",
    title          = "{Electron and photon performance measurements with the ATLAS detector using the 2015--2017 LHC proton--proton collision data}",
    journal        = "JINST",
    volume         = "14",
    year           = "2019",
    pages          = "P12006",
    doi            = "10.1088/1748-0221/14/12/P12006",
    reportNumber   = "CERN-EP-2019-145",
    eprint         = "1908.00005",
    archivePrefix  = "arXiv",
    primaryClass   = "hep-ex",
}

@Article{FTAG-2018-01,
    author         = "{ATLAS Collaboration}",
    title          = "{ATLAS \(b\)-jet identification performance and efficiency measurement with \(t\bar{t}\) events in \(pp\) collisions at \(\sqrt{s} = 13\,\text{TeV}\)}",
    journal        = "Eur. Phys. J. C",
    volume         = "79",
    year           = "2019",
    pages          = "970",
    doi            = "10.1140/epjc/s10052-019-7450-8",
    reportNumber   = "CERN-EP-2019-132",
    eprint         = "1907.05120",
    archivePrefix  = "arXiv",
    primaryClass   = "hep-ex",
}

@Article{JETM-2018-05,
    author         = "{ATLAS Collaboration}",
    title          = "{Jet energy scale and resolution measured in proton--proton collisions at \(\sqrt{s} = 13\,\text{TeV}\) with the ATLAS detector}",
    journal        = "Eur. Phys. J. C",
    volume         = "81",
    year           = "2020",
    pages          = "689",
    doi            = "10.1140/epjc/s10052-021-09402-3",
    reportNumber   = "CERN-EP-2020-083",
    eprint         = "2007.02645",
    archivePrefix  = "arXiv",
    primaryClass   = "hep-ex",
}

@Article{MUON-2018-03,
    author         = "{ATLAS Collaboration}",
    title          = "{Muon reconstruction and identification efficiency in ATLAS using the full Run~2 \(pp\) collision data set at \(\sqrt{s} = 13\,\text{TeV}\)}",
    journal        = "Eur. Phys. J. C",
    volume         = "81",
    year           = "2021",
    pages          = "578",
    doi            = "10.1140/epjc/s10052-021-09233-2",
    reportNumber   = "CERN-EP-2020-199",
    eprint         = "2012.00578",
    archivePrefix  = "arXiv",
    primaryClass   = "hep-ex",
}

@Article{PIX-2018-001,
    author         = "Abbott, B. and others",
    title          = "{Production and integration of the ATLAS Insertable B-Layer}",
    journal        = "JINST",
    volume         = "13",
    year           = "2018",
    pages          = "T05008",
    doi            = "10.1088/1748-0221/13/05/T05008",
    eprint         = "1803.00844",
    archivePrefix  = "arXiv",
    primaryClass   = "physics.ins-det",
}

@Article{SUSY-2018-09,
    author         = "{ATLAS Collaboration}",
    title          = "{Search for squarks and gluinos in final states with same-sign leptons and jets using \(139\,\text{fb}^{-1}\) of data collected with the ATLAS detector}",
    journal        = "JHEP",
    volume         = "06",
    year           = "2020",
    pages          = "046",
    doi            = "10.1007/JHEP06(2020)046",
    reportNumber   = "CERN-EP-2019-161",
    eprint         = "1909.08457",
    archivePrefix  = "arXiv",
    primaryClass   = "hep-ex",
}

@Article{TOPQ-2018-01,
    author         = "{ATLAS Collaboration}",
    title          = "{Observation of the associated production of a top quark and a \(Z\) boson in \(pp\) collisions at \(\sqrt{s} = 13\,\text{TeV}\) with the ATLAS detector}",
    journal        = "JHEP",
    volume         = "07",
    year           = "2020",
    pages          = "124",
    doi            = "10.1007/JHEP07(2020)124",
    reportNumber   = "CERN-EP-2019-273",
    eprint         = "2002.07546",
    archivePrefix  = "arXiv",
    primaryClass   = "hep-ex",
}

@Article{TOPQ-2018-11,
    author         = "{ATLAS Collaboration}",
    title          = "{Differential \(t\bar{t}\) cross-section measurements using boosted top quarks in the all-hadronic final state with \(139\,\text{fb}^{-1}\) of ATLAS data}",
    eprint = "2205.02817",
    archivePrefix = "arXiv",
    primaryClass = "hep-ex",
    reportNumber = "CERN-EP-2022-026",
    doi = "10.1007/JHEP04(2023)080",
    journal = "JHEP",
    volume = "04",
    pages = "080",
    year = "2023"
}

@Article{TRIG-2018-01,
    author         = "{ATLAS Collaboration}",
    title          = "{Performance of the ATLAS muon triggers in Run~2}",
    journal        = "JINST",
    volume         = "15",
    year           = "2020",
    pages          = "P09015",
    doi            = "10.1088/1748-0221/15/09/p09015",
    reportNumber   = "CERN-EP-2020-031",
    eprint         = "2004.13447",
    archivePrefix  = "arXiv",
    primaryClass   = "hep-ex",
}

@Article{TRIG-2018-05,
    author         = "{ATLAS Collaboration}",
    title          = "{Performance of electron and photon triggers in ATLAS during LHC Run~2}",
    journal        = "Eur. Phys. J. C",
    volume         = "80",
    year           = "2020",
    pages          = "47",
    doi            = "10.1140/epjc/s10052-019-7500-2",
    reportNumber   = "CERN-EP-2019-169",
    eprint         = "1909.00761",
    archivePrefix  = "arXiv",
    primaryClass   = "hep-ex",
}

@Article{TOPQ-2019-23,
    author         = "{ATLAS Collaboration}",
    title          = "{Measurements of differential cross-sections in top-quark pair events with a high transverse momentum top quark and limits on beyond the Standard Model contributions to top-quark pair production with the ATLAS detector at \(\sqrt{s} = 13\,\text{TeV}\)}",
    journal        = "JHEP",
    volume         = "06",
    year           = "2022",
    pages          = "063",
    doi            = "10.1007/JHEP06(2022)063",
    reportNumber   = "CERN-EP-2022-003",
    eprint         = "2202.12134",
    archivePrefix  = "arXiv",
    primaryClass   = "hep-ex",
}

@Article{TOPQ-2019-28,
    author         = "{ATLAS Collaboration}",
    title          = "{Measurement of the energy asymmetry in \(t\bar{t}j\) production at \(13\,\text{TeV}\) with the ATLAS experiment and interpretation in the SMEFT framework}",
    journal        = "Eur. Phys. J. C",
    volume         = "82",
    year           = "2022",
    pages          = "374",
    doi            = "10.1140/epjc/s10052-022-10101-w",
    reportNumber   = "CERN-EP-2021-181",
    eprint         = "2110.05453",
    archivePrefix  = "arXiv",
    primaryClass   = "hep-ex",
}

@Article{TOPQ-2020-06,
    author         = "{ATLAS Collaboration}",
    title          = "{Evidence for the charge asymmetry in \(pp \to t\bar{t}\) production at \(\sqrt{s} = 13\,\text{TeV}\) with the ATLAS detector}",
    year           = "2022",
    reportNumber   = "CERN preprint ID: CERN-EP-2022-166",
    eprint         = "2208.12095",
    archivePrefix  = "arXiv",
    primaryClass   = "hep-ex",
}

@Article{TOPQ-2020-10,
    author         = "{ATLAS Collaboration}",
    title          = "{Measurement of the \(t\bar{t}t\bar{t}\) production cross section in \(pp\) collisions at \(\sqrt{s} = 13\,\text{TeV}\) with the ATLAS detector}",
    journal        = "JHEP",
    volume         = "11",
    year           = "2021",
    pages          = "118",
    doi            = "10.1007/JHEP11(2021)118",
    reportNumber   = "CERN-EP-2021-075",
    eprint         = "2106.11683",
    archivePrefix  = "arXiv",
    primaryClass   = "hep-ex",
}

@Booklet{ATL-SOFT-PUB-2021-001,
    author         = "{ATLAS Collaboration}",
    title          = "{The ATLAS Collaboration Software and Firmware}",
    howpublished   = "{ATL-SOFT-PUB-2021-001}",
    url            = "https://cds.cern.ch/record/2767187",
    year           = "2021",
}

@Booklet{ATL-SOFT-PUB-2021-003,
    author         = "{ATLAS Collaboration}",
    title          = "{ATLAS Computing Acknowledgements}",
    howpublished   = "{ATL-SOFT-PUB-2021-003}",
    url            = "https://cds.cern.ch/record/2776662",
}

@Report{ATLAS-TDR-2010-19,
    author         = "{ATLAS Collaboration}",
    title          = "{ATLAS Insertable B-Layer Technical Design Report}",
    type           = "ATLAS-TDR-19; CERN-LHCC-2010-013",
    year           = "2010",
    url            = "https://cds.cern.ch/record/1291633",
    related        = "ATLAS-TDR-2010-19-addm",
    relatedstring  = "Addendum:",
}

@Article{CMS-TOP-17-014,
    author         = "{CMS Collaboration}",
    title          = "{Measurements of \(t\bar{t}\) differential cross sections in proton--proton collisions at \(\sqrt{s} = 13\,\text{TeV}\) using events containing two leptons}",
    journal        = "JHEP",
    volume         = "02",
    year           = "2019",
    pages          = "149",
    doi            = "10.1007/JHEP02(2019)149",
    reportNumber   = "CERN-EP-2018-252",
    eprint         = "1811.06625",
    archivePrefix  = "arXiv",
    primaryClass   = "hep-ex",
}

@Booklet{ATLAS-CONF-2014-018,
    author         = "{ATLAS Collaboration}",
    title          = "{Tagging and suppression of pileup jets with the ATLAS detector}",
    howpublished   = "{ATLAS-CONF-2014-018}",
    url            = "https://cds.cern.ch/record/1700870",
    year           = "2014",
}

@Booklet{ATLAS-CONF-2018-006,
    author         = "{ATLAS Collaboration}",
    title          = "{Calibration of light-flavour \(b\)-jet mistagging rates using ATLAS proton--proton collision data at \(\sqrt{s} = 13~\text{TeV}\)}",
    howpublished   = "{ATLAS-CONF-2018-006}",
    url            = "https://cds.cern.ch/record/2314418",
    year           = "2018",
}

@Booklet{ATLAS-CONF-2019-021,
    author         = "{ATLAS Collaboration}",
    title          = "{Luminosity determination in \(pp\) collisions at \(\sqrt{s} = 13\,\text{TeV}\) using the ATLAS detector at the LHC}",
    howpublished   = "{ATLAS-CONF-2019-021}",
    url            = "https://cds.cern.ch/record/2677054",
    year           = "2019",
}

@Booklet{ATL-PHYS-PUB-2014-021,
    author         = "{ATLAS Collaboration}",
    title          = "{ATLAS Pythia~8 tunes to \(7~\text{TeV}\) data}",
    howpublished   = "{ATL-PHYS-PUB-2014-021}",
    url            = "https://cds.cern.ch/record/1966419",
    year           = "2014",
}

@Booklet{ATL-PHYS-PUB-2016-017,
    author         = "{ATLAS Collaboration}",
    title          = "{The Pythia~8 A3 tune description of ATLAS minimum bias and inelastic measurements incorporating the Donnachie--Landshoff diffractive model}",
    howpublished   = "{ATL-PHYS-PUB-2016-017}",
    url            = "https://cds.cern.ch/record/2206965",
    year           = "2016",
}

@Booklet{ATL-PHYS-PUB-2016-020,
    author         = "{ATLAS Collaboration}",
    title          = "{Studies on top-quark Monte Carlo modelling for Top2016}",
    howpublished   = "{ATL-PHYS-PUB-2016-020}",
    url            = "https://cds.cern.ch/record/2216168",
    year           = "2016",
}

@Booklet{ATL-PHYS-PUB-2017-013,
    author         = "{ATLAS Collaboration}",
    title          = "{Optimisation and performance studies of the ATLAS \(b\)-tagging algorithms for the 2017-18 LHC run}",
    howpublished   = "{ATL-PHYS-PUB-2017-013}",
    url            = "https://cds.cern.ch/record/2273281",
    year           = "2017",
}
 
\clearpage
 
\begin{flushleft}
\hypersetup{urlcolor=black}
{\Large The ATLAS Collaboration}

\bigskip

\AtlasOrcid[0000-0002-6665-4934]{G.~Aad}$^\textrm{\scriptsize 102}$,
\AtlasOrcid[0000-0002-5888-2734]{B.~Abbott}$^\textrm{\scriptsize 120}$,
\AtlasOrcid[0000-0002-7248-3203]{D.C.~Abbott}$^\textrm{\scriptsize 103}$,
\AtlasOrcid[0000-0002-1002-1652]{K.~Abeling}$^\textrm{\scriptsize 55}$,
\AtlasOrcid[0000-0002-8496-9294]{S.H.~Abidi}$^\textrm{\scriptsize 29}$,
\AtlasOrcid[0000-0002-9987-2292]{A.~Aboulhorma}$^\textrm{\scriptsize 35e}$,
\AtlasOrcid[0000-0001-5329-6640]{H.~Abramowicz}$^\textrm{\scriptsize 151}$,
\AtlasOrcid[0000-0002-1599-2896]{H.~Abreu}$^\textrm{\scriptsize 150}$,
\AtlasOrcid[0000-0003-0403-3697]{Y.~Abulaiti}$^\textrm{\scriptsize 117}$,
\AtlasOrcid[0000-0003-0762-7204]{A.C.~Abusleme~Hoffman}$^\textrm{\scriptsize 137a}$
% ...
\bigskip
\\

% ...
$^{29}$Physics Department, Brookhaven National Laboratory, Upton NY; United States of America.\\
% ...
$^{35}$$^{(a)}$Facult\'e des Sciences Ain Chock, R\'eseau Universitaire de Physique des Hautes Energies - Universit\'e Hassan II, Casablanca;$^{(b)}$Facult\'{e} des Sciences, Universit\'{e} Ibn-Tofail, K\'{e}nitra;$^{(c)}$Facult\'e des Sciences Semlalia, Universit\'e Cadi Ayyad, LPHEA-Marrakech;$^{(d)}$LPMR, Facult\'e des Sciences, Universit\'e Mohamed Premier, Oujda;$^{(e)}$Facult\'e des sciences, Universit\'e Mohammed V, Rabat;$^{(f)}$Institute of Applied Physics, Mohammed VI Polytechnic University, Ben Guerir; Morocco.\\
% ...
$^{55}$II. Physikalisches Institut, Georg-August-Universit\"{a}t G\"ottingen, G\"ottingen; Germany.\\
% ...
$^{102}$CPPM, Aix-Marseille Universit\'e, CNRS/IN2P3, Marseille; France.\\
$^{103}$Department of Physics, University of Massachusetts, Amherst MA; United States of America.\\
% ...
$^{117}$Department of Physics, New York University, New York NY; United States of America.\\
% ...
$^{120}$Homer L. Dodge Department of Physics and Astronomy, University of Oklahoma, Norman OK; United States of America.\\
% ...
$^{137}$$^{(a)}$Departamento de F\'isica, Pontificia Universidad Cat\'olica de Chile, Santiago;$^{(b)}$Millennium Institute for Subatomic physics at high energy frontier (SAPHIR), Santiago;$^{(c)}$Instituto de Investigaci\'on Multidisciplinario en Ciencia y Tecnolog\'ia, y Departamento de F\'isica, Universidad de La Serena;$^{(d)}$Universidad Andres Bello, Department of Physics, Santiago;$^{(e)}$Instituto de Alta Investigaci\'on, Universidad de Tarapac\'a, Arica;$^{(f)}$Departamento de F\'isica, Universidad T\'ecnica Federico Santa Mar\'ia, Valpara\'iso; Chile.\\
% ...
$^{150}$Department of Physics, Technion, Israel Institute of Technology, Haifa; Israel.\\
$^{151}$Raymond and Beverly Sackler School of Physics and Astronomy, Tel Aviv University, Tel Aviv; Israel.\\
% ...
$^{a}$ Also Affiliated with an institute covered by a cooperation agreement with CERN.\\
$^{b}$ Also at An-Najah National University, Nablus; Palestine.\\
$^{c}$ Also at Borough of Manhattan Community College, City University of New York, New York NY; United States of America.\\
$^{d}$ Also at Bruno Kessler Foundation, Trento; Italy.\\
$^{e}$ Also at Center for High Energy Physics, Peking University; China.\\
$^{f}$ Also at Center for Interdisciplinary Research and Innovation (CIRI-AUTH), Thessaloniki ; Greece.\\
% ...

\end{flushleft}


\end{document}